\pdfoutput=1

\documentclass[final,5p,times,twocolumn,number]{elsarticle}

\journal{Computer Methods and Programs in Biomedicine}

\usepackage{amsmath}
\usepackage{amssymb}
\usepackage{amsfonts}

\usepackage{graphicx}
\usepackage{subcaption}
\usepackage{stfloats}
\usepackage{placeins}
\usepackage{balance}
\usepackage{booktabs}
\usepackage{threeparttable}
\usepackage{makecell}
\usepackage{multirow}
\usepackage{tabularx}
\usepackage{array}
\usepackage{siunitx}

\usepackage{algorithm}
\usepackage{algpseudocode}

\usepackage{xcolor}
\usepackage{colortbl}
\definecolor{bestcell}{RGB}{222,239,214}
\usepackage{url}

\usepackage[hidelinks]{hyperref}
\definecolor{linkblue}{RGB}{0,84,166}
\makeatletter
\def\ps@pprintTitle{%
  \let\@oddhead\@empty\let\@evenhead\@empty
  \let\@oddfoot\@empty\let\@evenfoot\@empty}
\makeatother

\graphicspath{{figures/}}

\newcommand{\RespGeomLib}{CT-RespGeomLib}

\AddToHook{cmd/finalMaketitle/after}{%
    \par\medskip
    \noindent\begin{minipage}{\textwidth}
        \captionsetup{type=figure}
        \centering
        \includegraphics[width=0.86\linewidth]{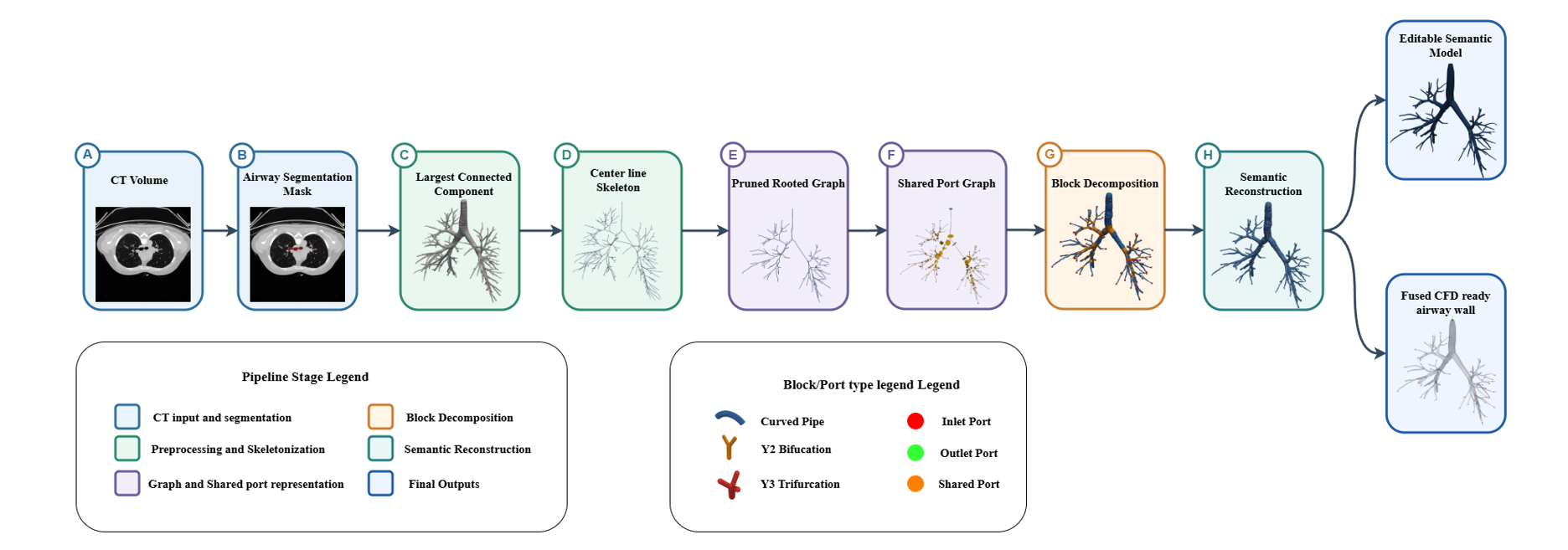}
        \caption{Overview of the CT-RespGeomLib reconstruction pipeline. A CT-derived airway mask is converted to a
        centreline--radius graph, rooted and pruned into an anatomical tree, decomposed into shared-port semantic
        blocks, and turned into two complementary outputs: an editable labelled block assembly and a fused single-wall
        surface with named ports for downstream meshing.}
        \label{fig:pipeline_overview}
    \end{minipage}
    \par\medskip
}

\begin{document}

\begin{frontmatter}

\title{CT-RespGeomLib: Deterministic reconstruction of CT airway masks into editable, port-matched lumen geometry}

\author[eee]{Nichula Wasalathilaka\corref{cor1}}
\ead{e20425@eng.pdn.ac.lk}
\cortext[cor1]{Corresponding author.}
\author[mech,jrdc]{Kevin Dawson}
\author[mech,swin]{Chinthaka Jacob}
\author[eee]{Mervyn Parakrama Ekanayake}
\author[eee]{Roshan Godaliyadda}

\affiliation[eee]{
    organization={Department of Electrical and Electronic Engineering, University of Peradeniya},
    city={Peradeniya},
    country={Sri Lanka}
}

\affiliation[mech]{
    organization={Department of Mechanical Engineering, University of Peradeniya},
    city={Peradeniya},
    country={Sri Lanka}
}

\affiliation[jrdc]{
    organization={China--Sri Lanka Joint Research and Demonstration Centre for Water Technology (JRDC), University of Peradeniya},
    city={Peradeniya},
    country={Sri Lanka}
}

\affiliation[swin]{
    organization={Department of Mechanical and Product Design Engineering, Swinburne University of Technology},
    city={Melbourne},
    state={Victoria},
    country={Australia}
}

\begin{abstract}
Patient-specific airway modelling requires geometrically consistent lumen surfaces with identifiable boundaries and support for controlled anatomical modifications. We present \RespGeomLib{}, a deterministic framework that converts CT airway masks into two complementary representations: an editable assembly of semantically labelled pipe and junction blocks connected through canonical shared ports, and a fused lumen surface reconstructed from tapered implicit primitives with validated port-to-boundary-loop correspondence. The parametric representation further enables localised modelling of airway stenosis and dilation without re-segmentation. Evaluated on 42 reference airway masks from ATM22 (15 cases) and AeroPath (27 cases), the framework produced 37 topologically accepted reconstructions, with a median port-to-loop distance of 0.123\,mm across 2174 matched ports. Reconstructed and reference equivalent radii showed strong agreement ($R^2=0.976$), with a mean bias of $-0.414$\,mm arising primarily from the circular, inscribed-radius representation. Controlled pathology modelling generated 40 variants across five patient-specific geometries, including 25--75\% stenosis and 50\% dilation at tracheal and main-bronchus locations. Prescribed radius changes were reproduced with a mean absolute ratio error of 0.0056, while the median 95th-percentile deviation outside the modified region was 0.111\,mm. Downstream applicability was demonstrated through volume meshing and an OpenFOAM simulation with individually prescribed model-based resistance conditions at 62 outlets. Although additional surface preparation may be required for numerical meshing, \RespGeomLib{} provides a reproducible framework for CT-derived airway reconstruction, quantitative boundary validation and controlled geometric pathology modelling. A project website is available at \textcolor{linkblue}{\url{https://nichula01.github.io/CT-Resp/}}.
	
\end{abstract}


\end{frontmatter}

\section{Introduction and Related Work}
\label{sec:introduction}
\label{sec:intro}

Patient-specific airway geometry is a central input to image-based
pulmonary modelling, supporting quantitative morphometry and
computational studies of airflow and particle
transport~\cite{lin2013multiscale,kleinstreuer2010airflow,burrowes2017image}. In such studies, computed flow
depends not only on the prescribed boundary conditions but also on
the geometry of the reconstructed
lumen~\cite{debacker2008flow,cherobin2018sensitivity}.
Recent multiscale respiratory simulations have further highlighted
the influence of airway surface processing and distal boundary
conditions on flow and particle transport
predictions~\cite{nguyen2025multiscale}.
Because a conducting-airway model contains many physically distinct
outlets, these boundaries must be individually identifiable when
branch-specific conditions are imposed. Consequently, converting
segmented airways into geometrically consistent surfaces with
identifiable inlet and outlet boundaries is an important step
between medical image analysis and downstream numerical modelling.

Large-scale airway segmentation studies, including ATM22 and
AeroPath~\cite{zhang2023multisite,stoverud2024aeropath}, have
improved the availability of annotated CT data spanning different
acquisition settings and pathological conditions. Recent
anatomy-aware segmentation approaches have also addressed
branch continuity and topological completeness
~\cite{wang2024airwaytopology}.
Nevertheless, a binary segmentation does not directly provide
an editable representation of individual airway branches or
explicitly labelled geometric interfaces. Direct isosurface
extraction, for example using marching cubes
~\cite{lorensen1987mc}, reproduces the voxel-defined boundary
but may require smoothing, trimming and surface repair before
volume meshing. Automated approaches have addressed parts of
this process: Lauria et al.~\cite{lauria2022airwaymesh} developed
a method for generating triangulated pulmonary airway meshes
from segmented CT images, while Bartol et
al.~\cite{bartol2025subjectspecific} presented a subject-specific
CT-to-CFD workflow incorporating automated inlet and outlet
processing, surface preparation and volume meshing. These studies
establish important routes from segmentation to numerical
simulation, while also motivating explicit assessment of the
geometric and topological properties of the resulting surfaces.

A complementary body of work represents airways through
centrelines, parametric primitives or procedural tree models.
Tawhai et al.~\cite{tawhai2004ct} fitted curved
centreline--diameter models to CT-derived airway trees and
extended their distal structure, while Miyawaki et
al.~\cite{miyawaki2017automatic} developed subject-specific
reconstruction from CT-segmented centrelines and surfaces,
including bifurcations and trifurcations. Centreline-based
methods have also been used for automated discretisation of
branched tubular domains~\cite{marchandise2013mesh}.
Procedural approaches provide complementary capabilities for
controlling airway structure, including volume-filling tree
growth~\cite{tawhai2000generation}, space-filling
three-dimensional tree models~\cite{kitaoka1999three}, the AVATREE
framework~\cite{nousias2020avatree} and
RespGeomLib~\cite{wasalathilaka2026respgeomlib}.
RespGeomLib, in particular, uses specification-driven
port-based assembly and implicit junction blending.
More recently, Spasov et al.~\cite{spasov2026bronchialtrees}
combined CT-derived proximal anatomy with stochastic growth
to construct extended bronchial trees. Related image-to-model
frameworks in vascular modelling have integrated centreline
analysis, surface editing and numerical mesh
generation~\cite{antiga2008framework}.

Beyond reconstructing existing anatomy, controlled geometric
modification is valuable for investigating the effects of
airway narrowing. Ding et al.~\cite{ding2023stenotic}
examined how stenotic geometry influences airflow in central
airway obstruction, while Michaud-Dorko et
al.~\cite{michauddorko2024stenosis} investigated progressively
increasing subglottal constriction in an upper-airway model.
Such studies demonstrate the utility of controlled anatomical
variations for respiratory-flow investigations. For
patient-specific reconstruction frameworks, an additional
requirement is the ability to impose prescribed geometric
changes locally and verify their realised severity and
spatial extent independently of a subsequent flow simulation.

The present work addresses the relationship between an editable
airway representation and the physical surface reconstructed
from it. Existing methods provide important capabilities in
CT-based reconstruction, automated outlet preparation,
procedural geometry generation and downstream meshing.
However, the correspondence between a structured branch-level
representation and the boundary interfaces realised on a
separately reconstructed lumen surface remains a distinct
validation problem. In RespGeomLib, ports are defined by the
geometric specification and are therefore consistent within
the assembled representation. For CT-derived geometry,
branch structure and local dimensions must first be recovered
from image data, and the appearance of each intended external
port as a distinct, correctly located boundary loop on the
final surface must be established independently.

We present CT-RespGeomLib, a deterministic framework that
converts CT airway masks into two complementary geometric
outputs. A rooted centreline--radius graph is decomposed into
an editable assembly of labelled pipe and junction blocks
connected through shared canonical ports. A separate fused
lumen surface is reconstructed from graph-derived tapered
implicit primitives, opened at the physical inlet and outlets,
and evaluated using connectivity, edge-manifoldness and
one-to-one port--loop correspondence. The same representation
supports controlled, branch-local radius modifications for
modelling stenosis and dilation without re-segmenting the
original CT image. The framework is evaluated on 42 reference
airway masks from ATM22 and AeroPath through geometric
fidelity measurements, baseline comparisons, parameter
sensitivity analysis and 40 controlled pathology variants.
Downstream applicability is further examined through
volume meshing and an OpenFOAM flow simulation with
individual model-based outlet resistance conditions.

The principal contributions are:
\begin{itemize}
	
	\item A deterministic CT-mask-to-geometry pipeline that
	produces an editable semantic airway assembly and a
	separately reconstructed fused lumen surface in physical
	coordinates.
	
	\item A canonical shared-port representation together
	with explicit geometric validation of one-to-one
	correspondence between intended external ports and
	realised boundary loops on the fused surface.
	
	\item Controlled, branch-local modelling of stenosis
	and dilation in CT-derived airway geometries, with
	quantitative evaluation of prescribed versus realised
	severity, spatial locality and boundary preservation.
	
	\item A multi-cohort evaluation of reconstruction
	acceptance, geometric fidelity, baseline methods and
	parameter sensitivity, complemented by a downstream
	volume-meshing and OpenFOAM demonstration with
	individually prescribed outlet boundary conditions.
	
\end{itemize}

\providecommand{\Gr}{\mathcal{G}}
\providecommand{\Po}{\mathcal{P}}
\providecommand{\Bl}{\mathcal{B}}
\providecommand{\Kc}{\mathcal{K}}
\providecommand{\Mclean}{M_{\mathrm{clean}}}
\providecommand{\xworld}{\mathbf{x}_{\mathrm{world}}}
\providecommand{\CTRG}{CT\mbox{-}RespGeomLib}
\providecommand{\SM}[1]{\hyperref[sec:sm#1]{Supplementary Methods~SM#1}}

\section{Methods}
\label{sec:methods}

\subsection{Framework overview and datasets}
\label{sec:overview}
\label{sec:data}

\begin{figure*}[!t]
	\centering
	\includegraphics[width=\textwidth]{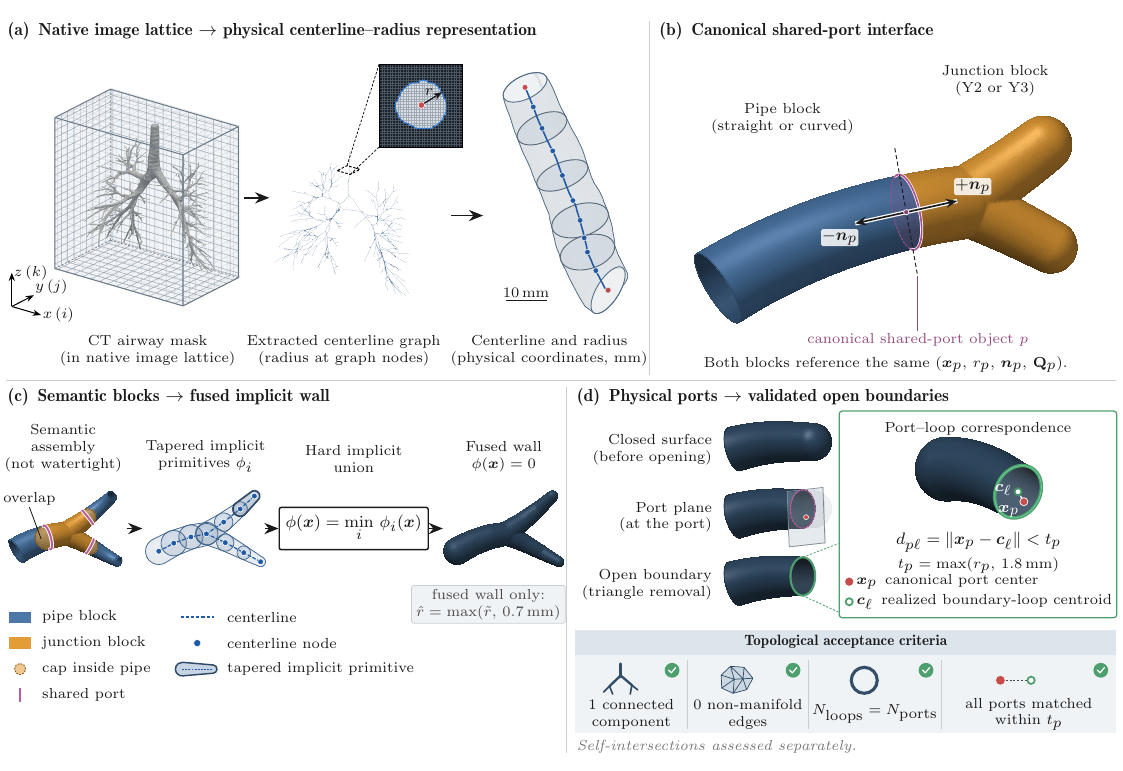}
	\caption{Four implementation details of the CT-RespGeomLib method.
		(a)~Voxel topology is analysed on the native anisotropic image lattice,
		while centrelines, radii, ports and surfaces are mapped to physical world
		coordinates in millimetres. (b)~Adjacent blocks reference the same shared
		port object, so interface centres, radii, frames and opposing normals are
		coincident by construction. (c)~Tapered implicit primitives
		are combined by a hard implicit union to form the fused wall.
		(d)~The closed surface is opened at physical ports, and the realised
		boundary loops are matched to the canonical ports under the topological
		acceptance criteria. The panels correspond to Eq.~\eqref{eq:affine} (a), Eq.~\eqref{eq:frame} (b),
		Eqs.~\eqref{eq:capsule-t}--\eqref{eq:union} (c) and Eq.~\eqref{eq:tolerance} (d).}
	\label{fig:method_panels}
\end{figure*}

\CTRG{} takes a binary CT airway mask and generates two complementary outputs: a semantically labelled assembly of editable airway blocks and a fused lumen surface for downstream geometric processing. Both use the same centreline--radius representation, although smoothing and radius constraints mean that their realised surfaces are not identical. For an input mask $M$, the editable semantic representation is defined by
\begin{equation}
	\mathcal{A}=(\Gr,\Po,\Bl),
	\label{eq:triple}
\end{equation}
where $\Gr=(V,E)$ is a rooted centreline--radius graph, $\Po$ contains canonical geometric ports, and $\Bl$ contains straight and curved pipes and Y2, Y3 and general junction blocks. Graph operations are performed on the native voxel lattice; geometry and outputs are expressed in millimetres. Figure~\ref{fig:pipeline_overview} and Algorithm~\ref{alg:ctrg} summarise the procedure, and Fig.~\ref{fig:method_panels} illustrates four implementation details.

The evaluation used 42 reference airway masks: 15 training cases from the ATM22 dataset~\cite{zhang2023multisite} and all 27 AeroPath cases~\cite{stoverud2024aeropath}. ATM22 provides multi-site chest CT airway annotations, while AeroPath includes contrast-enhanced CT angiography acquired in patients investigated for pulmonary disease. The published reference masks, rather than predictions of an additional segmentation model, were supplied to the reconstruction pipeline. Case selection and acquisition characteristics are documented in \SM{1}.

\begin{algorithm}[t]
	\caption{CT-RespGeomLib reconstruction pipeline.}
	\label{alg:ctrg}
	\label{alg:ct_respgeomlib}
	\begin{algorithmic}[1]
		\Require CT airway mask $M$ with origin, spacing and direction matrix
		\Ensure Semantic block assembly, fused surface $S$, port manifest and validation report
		\State $\Mclean \gets$ largest 18-connected foreground component; crop with eight-voxel margin
		\State Thin $\Mclean$; compute spacing-aware distance-transform radii
		\State Build $\Gr$ from 26-connected skeleton chains; root, prune and cap generations
		\State Define canonical ports $\Po$ and decompose $\Gr$ into semantic blocks $\Bl$
		\State Resample and smooth graph edges; restore endpoint and port parameters
		\State Construct labelled pipe and junction primitives from the port-linked blocks
		\State Apply fused-wall radius floor; construct tapered implicit segments
		\State Evaluate the hard-minimum field on the selected physical-space grid
		\State Extract zero isosurface $S$; open the physical inlet and outlets
		\State Smooth and repair $S$ while retaining all port-bearing components
		\State Extract boundary loops and assign external ports one-to-one within tolerance
		\State Report connectivity, manifold edges, port--loop matches and acceptance
	\end{algorithmic}
\end{algorithm}

\begin{table}[t]
	\centering
	\caption{Principal reconstruction settings. The complete parameter list, with the purpose of each setting, is given in Supplementary Table~\ref{tab:params-full}.}
	\label{tab:params}
	\small
	\setlength{\tabcolsep}{4pt}
	\begin{tabular}{@{}ll@{}}
		\toprule
		Setting & Value \\
		\midrule
		Mask component connectivity & 18 \\
		Skeleton graph connectivity & 26 \\
		Mask crop margin & 8 voxels \\
		Terminal spur length / mean radius & $8.0$ / $0.75\,\mathrm{mm}$ \\
		Maximum generation & 8 \\
		Junction cut factor $\alpha$ & 2.0 \\
		Curved-pipe circumferential samples & 96 \\
		Fused radius floor $\rho$ & $0.7\,\mathrm{mm}$ \\
		Implicit nearest-segment count $K$ & 24 \\
		Grid spacing / point budget & $0.40$--$1.50\,\mathrm{mm}$ / $6\times10^7$ \\
		Taubin smoothing & 40 iterations; pass band 0.05 \\
		Hole-fill limit & $0.56\,\mathrm{mm}$ \\
		Port-centroid tolerance & $\max(r_p,1.8\,\mathrm{mm})$ \\
		\bottomrule
	\end{tabular}
\end{table}

\subsection{CT preprocessing and centreline--radius graph construction}
\label{sec:coords}
\label{sec:graph}
\label{sec:pruning}

NIfTI airway annotations are loaded with SimpleITK~\cite{lowekamp2013simpleitk}, treating strictly positive labels as foreground. Connectivity and thinning operate in array order $(z,y,x)$; physical geometry uses the ITK index $\mathbf{u}=(i,j,k)$ and image affine
\begin{equation}
	\xworld=\mathbf{o}+D(\mathbf{u}\odot\mathbf{s}),
	\label{eq:affine}
\end{equation}
where $\mathbf{o}$, $D$ and $\mathbf{s}$ denote origin, direction matrix and ITK-order voxel spacing, respectively, and $\odot$ is elementwise multiplication. The native mask is not resampled; exported coordinates follow ITK LPS convention.

The largest 18-connected foreground component is retained,
\begin{equation}
	\Mclean=\operatorname*{arg\,max}_{C_i\in\Kc_{18}(M)}|C_i|,
	\label{eq:largestcc}
\end{equation}
where $\Kc_{18}(M)$ denotes the set of 18-connected components. The retained mask is cropped with an eight-voxel margin; no morphological closing, dilation or hole filling is performed. A one-voxel-wide skeleton is obtained by three-dimensional thinning~\cite{lee1994thinning}. Skeleton radii are estimated from the anisotropic Euclidean distance transform,
\begin{equation}
	r(\mathbf{u})=\min_{\mathbf{y}\in\mathcal{N}(\Mclean)}
	\bigl\|(\mathbf{u}-\mathbf{y})\odot\mathbf{s}\bigr\|_2,
	\label{eq:edt}
\end{equation}
where $\mathcal{N}(\Mclean)$ denotes the background voxel centres considered by the distance transform. Distances are in millimetres and are not corrected by half a voxel; consequently, they provide voxel-centre-based estimates of the local inscribed radius.

The skeleton is traced using 26-neighbour connectivity. Adjacent critical voxels (degree other than two) are clustered into graph nodes, while intervening degree-two chains form edges. For each node pair the shortest traced chain is retained; chains shorter than three voxels are discarded. Each edge $e$ stores an ordered centreline $C_e$, sampled radii $r_e$ and physical arc length $\ell_e$.

The root maximises $(r_{\max},r_{\mathrm{mean}},z)$ lexicographically across endpoints. A shortest-path tree weighted by arc length then establishes parent--child orientation. Terminal edges shorter than $8.0\,\mathrm{mm}$ or with mean radius below $0.75\,\mathrm{mm}$ are pruned iteratively, except for the root edge. Degree-two pass-through nodes are collapsed, and the tree is truncated after eight branch-node generations, with the trachea assigned generation zero. Further details, including how often each step changed the input, are given in \SM{2}.

\subsection{Shared-port representation and semantic block assembly}
\label{sec:ports}
\label{sec:blocks}

Each port $p\in\Po$ stores a centre $\mathbf{x}_p$, unit normal $\mathbf{n}_p$, radius $r_p$, local frame $Q_p\in SO(3)$, type and identifier. Port types comprise physical inlets and outlets and internal junction-parent and junction-child cuts. External normals follow the centreline end tangents; junction-cut normals follow the chord from the junction node to the cut location. The frame is obtained by Gram--Schmidt orthogonalisation:
\begin{equation}
	\begin{aligned}
		\mathbf{e}_3&=\frac{\mathbf{n}_p}{\|\mathbf{n}_p\|_2},\\
		\mathbf{e}_1&=\frac{\mathbf{a}-(\mathbf{a}\cdot\mathbf{e}_3)\mathbf{e}_3}
		{\|\mathbf{a}-(\mathbf{a}\cdot\mathbf{e}_3)\mathbf{e}_3\|_2},\\
		\mathbf{e}_2&=\mathbf{e}_3\times\mathbf{e}_1,\qquad
		Q_p=[\,\mathbf{e}_1\ \mathbf{e}_2\ \mathbf{e}_3\,],
	\end{aligned}
	\label{eq:frame}
\end{equation}
with $\mathbf{a}=\mathbf{e}_x$, replaced by $\mathbf{e}_y$ when $|\mathbf{e}_x\cdot\mathbf{e}_3|>0.95$. Port radii are interpolated from the corresponding edge profiles.

Each canonical port is stored once, and neighbouring blocks reference its identifier rather than independent interface parameters. The canonical centre, radius and frame are therefore common to both blocks, with opposite block-local outward normals. Numerical assertions check centre discrepancies below $10^{-4}\,\mathrm{mm}$ and a normal dot product below $-0.95$ for paired interfaces. This representation-level consistency does not itself guarantee that a distinct external boundary loop appears on the subsequently fused surface; that property is validated separately (Section~\ref{sec:acceptance}).

Graph branches are divided into pipes and junction regions using a radius-dependent cut distance
\begin{equation}
	\delta=\min\!\left(\max(\alpha r_{\mathrm{node}},\varepsilon),\;0.4\ell_e\right),
	\qquad\alpha=2.0,\quad\varepsilon=10^{-3}\,\mathrm{mm},
	\label{eq:cut}
\end{equation}
where $r_{\mathrm{node}}$ is the node radius and $\ell_e$ the incident edge length. Junctions are classified by daughter count (two: Y2; three: Y3; four or more: general). A pipe is designated curved when its arc-to-chord ratio exceeds 1.03 or its maximum chord deviation exceeds 5\% of its length; otherwise it is straight.

Curved pipes are swept along smoothed centreline--radius profiles using rotation-minimising frames~\cite{wang2008rmf}; straight pipes use tapered chord-aligned cylinders. Y2 and Y3 junctions are locally contoured implicit unions of parent and daughter tube fields. Their rounded ends overlap neighbouring pipes because the junction surfaces are not trimmed at the canonical cut planes. The output is a labelled VTK PolyData assembly of overlapping primitives, not a watertight meshing surface. General junctions preserve their graph and port metadata but are not explicitly realised as semantic primitives; their branches remain represented in the fused field. The block census of the cohort and the primitive construction are reported in \SM{3}.

\subsection{Implicit lumen-surface reconstruction}
\label{sec:fused}

The fused wall is reconstructed from graph-derived implicit primitives, not by stitching semantic meshes. Graph edges are resampled in arc length, then centreline and radius profiles are Gaussian-smoothed with characteristic widths of approximately $0.08\ell_e$ and $0.18\ell_e$, respectively. Over the terminal 20\% of each edge, the smoothed profiles are blended towards their port values and both endpoints are restored exactly. For the fused wall only, the radius is constrained by
\begin{equation}
	\hat r_e(\xi)=\max\!\left(\tilde r_e(\xi),\rho\right),
	\qquad \rho=0.7\,\mathrm{mm},
	\label{eq:floor}
\end{equation}
where $\xi$ denotes arc length and $\tilde r_e$ the smoothed profile. The radius floor enlarges lumens smaller than $\rho$ and is evaluated as a source of geometric deviation.

A tapered segment with endpoints $\mathbf{a},\mathbf{b}$ and end radii $r_a,r_b$ defines
\begin{align}
	\mathbf{v}&=\mathbf{b}-\mathbf{a},\qquad
	t=\operatorname{clip}\!\left(
	\frac{(\mathbf{x}-\mathbf{a})\cdot\mathbf{v}}{\mathbf{v}\cdot\mathbf{v}},0,1\right),
	\label{eq:capsule-t}\\
	\varphi_{\mathrm{seg}}(\mathbf{x})&=
	\|\mathbf{x}-(\mathbf{a}+t\mathbf{v})\|_2
	-\left[r_a+t(r_b-r_a)\right].
	\label{eq:capsule-phi}
\end{align}
For constant radius, $\varphi_{\mathrm{seg}}$ is the exact signed-distance function of a capsule; for varying radius it is an implicit approximation. The combined field is the hard minimum
\begin{equation}
	\varphi(\mathbf{x})=\min_{i\in N_K(\mathbf{x})}
	\varphi_{\mathrm{seg},i}(\mathbf{x}),\qquad K=24,
	\label{eq:union}
\end{equation}
where $N_K(\mathbf{x})$ comprises the segments with the nearest midpoints. This avoids additional smooth-union blending but may cause neighbouring branches to influence one another's wall where their primitives overlap.

A midpoint $k$-d tree supports restricted-band evaluation. The field is sampled on a physical-space Cartesian grid whose spacing $h$ is the finest admissible value from a fixed ladder spanning $0.40$--$1.50\,\mathrm{mm}$ under a $6\times10^7$-point budget. The zero isosurface $S=\{\mathbf{x}:\varphi(\mathbf{x})=0\}$ is extracted with VTK~\cite{schroeder2006vtk}; reconstruction fails if no admissible grid spacing satisfies the budget. The consequences of the smoothing scale, the narrow band and the grid selection are described in \SM{4}.

\subsection{Boundary preservation and geometric validation}
\label{sec:cleanup}
\label{sec:acceptance}

The initially closed fused surface is opened at external ports by removing triangles whose centroids lie beyond the corresponding port plane, $(\mathbf{c}-\mathbf{x}_p)\cdot\mathbf{n}_p>0$, within a port-specific cap region. Connected components carrying expected ports are retained. Boundary-preserving Taubin smoothing~\cite{taubin1995curve} (40 iterations, pass band 0.05), degenerate-triangle removal, iterative non-manifold-edge repair and size-limited hole filling are then applied. Component selection is repeated after repair; implementation thresholds are listed in Table~\ref{tab:params} and Supplementary Table~\ref{tab:params-full}.

A reconstruction is \emph{topologically accepted} when (i) the surface has one connected component, (ii) no non-manifold edges remain, (iii) the number of open boundary loops equals $1+N_{\mathrm{outlets}}$, and (iv) every physical port is matched to a distinct boundary loop. Internal junction cuts are not expected to appear as external loops. Boundary loops are formed from edges incident to one triangle and summarised by their centroid and mean radius. One-to-one assignment minimises port-to-loop centroid distance, subject to
\begin{equation}
	t_p=\max\!\left(1.0\,r_p,\;1.8\,\mathrm{mm}\right).
	\label{eq:tolerance}
\end{equation}
Assignments outside tolerance are rejected; unmatched ports and surplus loops are recorded individually. Port-to-loop radius error is measured but is not an acceptance condition because fused radii may be floored. Port-distance statistics include all matched external ports from the 42 attempted reconstructions, including cases that failed overall acceptance. Self-intersecting triangle pairs are counted independently: topological acceptance does not establish self-intersection freedom or guarantee direct volume meshability. The rationale for the clipping, component-selection and tolerance choices is given in \SM{5}.

\subsection{Controlled pathology editing}
\label{sec:pathology}

Controlled geometric stenosis and dilation are generated by altering the smoothed radius profile of a selected branch while leaving its centreline and graph unchanged. With normalised arc length $u=s/L$, the edited radius is
\begin{equation}
	r_{\mathrm{edit}}(u)=r_0(u)\left[1+\alpha\exp\!\left(
	-\frac{(u-c)^2}{2w^2}\right)\right],
	\label{eq:lesion}
\end{equation}
with $c=0.45$ and $w=0.10$, where $r_0$ is the healthy smoothed radius profile. Stenosis severities use $\alpha=-0.25,-0.50,-0.75$ and dilation uses $\alpha=0.50$, defining prescribed central radius ratios of $0.75$, $0.50$, $0.25$ and $1.50$. Branch endpoints are restored and the edited geometry is regenerated with the fixed production settings. The perturbations are geometric rather than disease-mechanical models.

Five healthy cases were modified at the trachea and at a main-bronchus branch selected by tracheal-chord alignment. Four edits at each site produced 40 variants. Realised severity was measured on matched healthy and edited surface cross-sections using the equivalent radius $r_{\mathrm{eq}}=\sqrt{A/\pi}$, with the ratio evaluated at $u=c$. Locality was evaluated outside the prespecified lesion interval $|u-c|\leq2.5w$, excluding fixed endpoint and junction margins. Each variant was independently checked against the acceptance criteria; detailed measurement definitions are reported in \SM{6}.

\subsection{Experimental evaluation}
\label{sec:evaluation}

Five strategies were compared on the same 42 cases: B1, direct marching cubes~\cite{lorensen1987mc}; B2, Taubin-smoothed B1~\cite{taubin1995curve}; B3, independently swept graph-edge tubes without union or junction trimming; B4, the labelled semantic assembly; and B5, the fused wall. Connected components, non-manifold edges, boundary-loop counts and self-intersecting triangle pairs were measured using a common evaluation pipeline. Port--loop correspondence was evaluated only for B5: B1--B3 have no canonical ports, whereas B4 stores ports in an overlapping semantic assembly whose internal interfaces are not realised external boundary loops.

Surface-to-mask agreement was evaluated from the unsigned distance of sampled mesh vertices to their nearest mask-boundary voxel centres, reporting median and 95th-percentile distances and the fraction of sampled vertices inside the mask. Lumen calibre was compared at $1.0\,\mathrm{mm}$ intervals along the raw centrelines, excluding junction neighbourhoods within twice the local radius. A common 180-ray, star-shaped cross-sectional estimator was applied to both mask and reconstructed surface; plane normals were estimated from a local principal direction of the raw centreline. Equivalent-radius agreement was summarised by $R^2$, root-mean-square error, mean bias and repeated-measurement Bland--Altman limits of agreement~\cite{blandaltman2007}. Case-level summaries were retained to account for within-case station clustering.

Branch angles, chord lengths, mean central-half diameters and Strahler and Horsfield orders were measured on the pruned trees. Ordering ratios used least-squares fits on $\log_{10}$ morphometric quantities against order, excluding the highest order and cases with fewer than four Strahler orders. Potential interaction between adjacent implicit branches was assessed by stratifying cross-sections according to the proximity of another branch's floored wall. The radius floor was swept over $\{0.5,0.6,0.7,0.8,0.9\}\,\mathrm{mm}$ and grid spacing over $\{0.40,0.45,0.50,0.60,0.80,1.00\}\,\mathrm{mm}$; all remaining settings were fixed. Runtime and peak resident memory were evaluated over three repetitions per case. Correlations and the comparison of self-intersection counts are exploratory. Estimator choices and the statistical treatment are detailed in \SM{7}.

\subsection{Downstream meshing and OpenFOAM demonstration}
\label{sec:cfd}
\label{sec:cfd-method}

One topologically accepted reconstruction was transferred to a volume-meshing and flow-simulation workflow. The fused STL surface was closed with caps at its inlet and 65 outlets, meshed with the Robust (Octree) tetrahedral method in ANSYS ICEM CFD 2023~R1, and imported into OpenFOAM v1912~\cite{weller1998foam}. The resulting mesh contained 656\,650 tetrahedral cells and 148\,083 nodes, without prism layers. Openings were separated into solver patches by connected surface regions; three outlets below $0.7\,\mathrm{mm}$ equivalent diameter were assigned to the wall, leaving 62 modelled outlets. The patch correspondence was generated from connected openings rather than directly from the exported canonical port identifiers. Mesh topology and quality were examined using \texttt{checkMesh}; no mesh-independence study was performed.

Air was treated as incompressible and Newtonian ($\rho_f=1.204\,\mathrm{kg\,m^{-3}}$, $\mu=1.825\times10^{-5}\,\mathrm{Pa\,s}$). A steady laminar simulation used a prescribed inlet flow of $0.1\,\mathrm{L\,s^{-1}}$ (inlet Reynolds number about 960) with a smooth wall-vanishing velocity profile, stationary no-slip walls, and model-derived resistance conditions at each outlet:
\begin{equation}
	p_k=R_kQ_k,
	\label{eq:outlet-resistance}
\end{equation}
where $Q_k$ is the simulated outlet flow and $R_k$ is estimated from a diameter-scaled symmetric Weibel subtree~\cite{weibel1963morphometry}. For an airway of generation $j$, diameter $D_j$ and length $L_j$, scaled by $s=D_k/D_g$, the individual tube resistance is
\begin{equation}
	R_j=\frac{128\mu\,s L_j}{\pi(sD_j)^4}.
	\label{eq:weibel-resistance}
\end{equation}
Series and parallel combinations extend the unresolved distal tree to generation 16. These are anatomy-scaled model resistances, not patient-measured peripheral resistances.

The steady equations were solved with \texttt{simpleFoam}, SIMPLEC pressure--velocity coupling~\cite{vandoormaal1984simplec} and bounded second-order convection. Outlet pressures were updated every 20 solver iterations from the computed patch flows. The stopping criteria required the maximum outlet-resistance mismatch to remain below 1\% and the largest outlet-flow-share change below 0.01 percentage points over two successive updates; equation residuals, mass balance and outlet backflow were checked separately. The cross-sections S1--S3, the split of flow between the lungs and the resistance-only reference share used in Section~\ref{sec:res-cfd} are defined in \SM{8}, together with the full mesh settings, inlet profile, resistance assignment and solver controls. This single-flow, single-mesh study demonstrates workflow integration and does not validate physiological airflow predictions.

\subsection{Implementation and reproducibility}
\label{sec:implementation}

The geometry pipeline uses Python 3.10, NumPy~\cite{harris2020numpy}, SciPy~\cite{virtanen2020scipy}, scikit-image~\cite{vanderwalt2014scikitimage}, SimpleITK and VTK through PyVista~\cite{sullivan2019pyvista}. It exports the semantic assembly and fused wall as VTK PolyData and STL, along with CSV/JSON port manifests recording identifiers, positions, normals, radii and structural boundary labels. Graph metadata, block assignments and validation reports are saved separately (\SM{9}). Three cases were reconstructed twice in separate processes and produced identical mesh topology and vertex coordinates under the tested environment. Timings were measured with eight pinned threads on an Intel Xeon Gold 5218R CPU. Code availability is stated at the end of the article.

\section{Results}
\label{sec:results}

\subsection{Reconstruction success and failure}
\label{sec:res-acceptance}

The pipeline was run on all 42 cases at the production settings of
Table~\ref{tab:params}. The automatically selected grid spacing ranged from 0.40 to 0.70\,mm.

\begin{figure*}[!t]
\centering
\includegraphics[width=\textwidth]{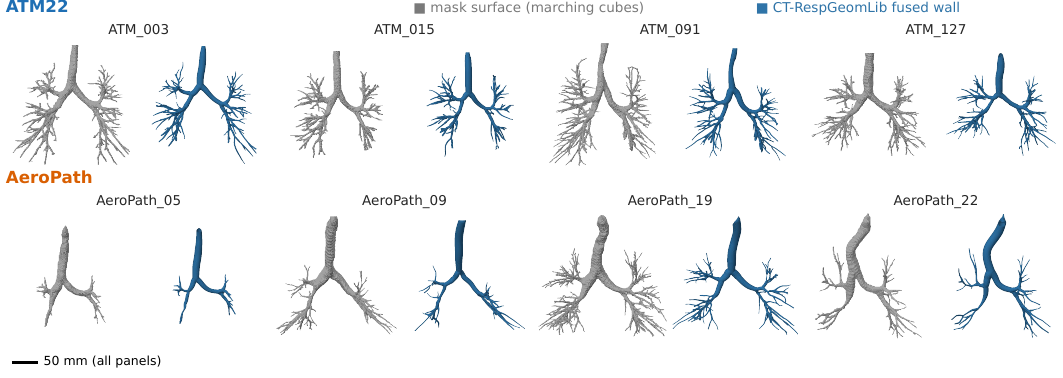}
\caption{Reference masks and reconstructions for four accepted cases from each cohort, front view. In each pair the
marching-cubes surface of the reference mask (grey) is shown beside the \RespGeomLib{} fused wall (blue); all panels
share one scale.}
\label{fig:qualitative}
\end{figure*}

Thirty-seven of the 42 reconstructions met every topological acceptance criterion of
Section~\ref{sec:acceptance}: 14 of 15 ATM22 cases and 23 of 27 AeroPath cases
(Table~\ref{tab:cohort}). Every accepted wall was a single connected component
with zero non-manifold edges, its number of open boundary loops equalled the
expected port count, and every port was matched to a distinct loop within tolerance. Figure~\ref{fig:qualitative} compares
reference-mask surfaces with the corresponding reconstructions;
\hyperref[sec:supplementary-results-1]{Supplementary Results~1} shows all
accepted cases and all five failures, and Supplementary Fig.~\ref{fig:s-percase} summarises every case.

The two cohorts differ in scale. Accepted ATM22 reconstructions contained
$76.2 \pm 28.0$ outlets against $34.6 \pm 17.4$ for AeroPath, whose annotated trees are shorter. The generation-8
cap was reached in 34 of 42 cases; in the remaining eight the segmented tree ended at generation~6 or~7.

\begin{figure*}[!t]
\centering
\includegraphics[width=\textwidth]{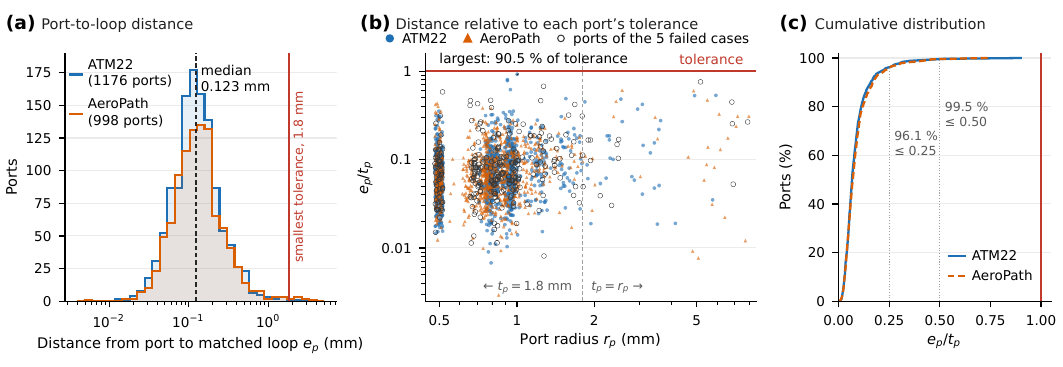}
\caption{Agreement between the canonical ports and the boundary loops of the fused wall, over all 2174 matched
external ports of the 42 cases. (a)~Distance $e_p$ from each port centre to the centroid of its matched loop, by
cohort. (b)~The same distance relative to the port's tolerance $t_p$ (Eq.~\eqref{eq:tolerance}) against port radius;
open circles are ports of the five cases that were not accepted. Below $r_p=1.8$\,mm the tolerance is the fixed
1.8\,mm, above it the port radius. (c)~Cumulative distribution of $e_p/t_p$. No port reaches its tolerance; the
largest ratio is 0.905.}
\label{fig:port-accuracy}
\end{figure*}

\paragraph{Port--loop agreement}
Port statistics include every matched external port of the 42 attempted reconstructions, including those of the
five rejected cases (Section~\ref{sec:acceptance}). Across these 2174 ports the median port-centre error was
0.123\,mm, with a 90th percentile of 0.304\,mm; 1.15\,\% of ports exceeded 1.0\,mm (Fig.~\ref{fig:port-accuracy}).
No matched port exceeded its tolerance; the largest error was 90.5\,\% of the port's tolerance. In the five rejected
cases every matched port lay within 74\,\% of its tolerance, and the rejections arose from surface topology: a
surplus boundary loop in four cases and, in AeroPath\_17, one port without a loop of its own (see \emph{Failure
analysis}).

Shared-port coincidence in the semantic assembly held to numerical precision (maximum centre discrepancy
$<10^{-14}$\,mm, mean interface dot product $-1.0000$ in every case), as required by construction
(Section~\ref{sec:ports}).

\paragraph{Determinism}
Three cases spanning both cohorts and the range of tree sizes were each
reconstructed twice in separate processes. Both runs produced identical point,
cell, component, non-manifold-edge and boundary-loop counts and a maximum per-vertex displacement of zero, on one
machine and thread configuration.

\paragraph{Self-intersections}
Topological acceptance does not imply a surface free of self-intersecting triangle pairs, which were counted
separately. Only 2 of the 42 fused walls had none; the median count was 5 (IQR 3--7, range 0--17). Counts increased
with tree size (Pearson $r=0.59$ against graph edges, $r=0.65$ against mesh faces; Supplementary
Fig.~\ref{fig:s-cost}d); the weaker correlation with grid spacing ($r=0.41$) is confounded, because spacing is selected
from tree size under the point budget. Medians were 5 in both accepted and rejected cases (Mann--Whitney $p=0.58$,
exploratory). An accepted wall may therefore still require self-intersection repair before volume meshing; the counts
are reported without repair.

\paragraph{Failure analysis}
The five rejections fall into two mechanisms.

In AeroPath\_17 one outlet port had no corresponding boundary loop. Its nearest
loop lay 1.36\,mm away but already served a neighbouring port 2.47\,mm from it:
the two branches run close enough that their capsule fields merge before
clipping, so the solid does not separate into two tubes and no per-port cut can
open them independently. This is the envelope behaviour of the hard union
(Section~\ref{sec:fused}) occurring at a port.

The other four cases each carry one or two surplus boundary loops --- openings
in the wall corresponding to no expected port --- while matching every expected
port correctly. Tracing these openings through each stage of
Section~\ref{sec:cleanup} shows that all of them first appear at the clipping
step and are untouched by smoothing, degenerate-triangle removal, repair and
hole filling; evaluating the field on a fine grid at each location confirms
that $\varphi$ changes sign there, so the implicit solid is continuous and the
opening is introduced by triangle removal rather than by the field. In three of
the four, the surplus loop accompanies a second connected component. No single mechanism common to all four was
identified.

\begin{table}[t]
\centering
\footnotesize
\setlength{\tabcolsep}{4pt}
\caption{Reconstruction outcome by cohort. Tree-size and topology rows are mean\,$\pm$\,SD over accepted cases;
port rows include all matched ports of all attempted cases; self-intersection medians are over all cases.}
\label{tab:cohort}
\begin{tabular}{lrr}
\toprule
Metric & ATM22 & AeroPath \\
\midrule
Cases attempted                      & 15 & 27 \\
Accepted                             & 14 & 23 \\
\midrule
Mean outlets, accepted cases         & $76.2 \pm 28.0$ & $34.6 \pm 17.4$ \\
Mean blocks, accepted cases          & $222 \pm 83$ & $100 \pm 50$ \\
Mean graph edges, accepted cases     & $149 \pm 55$ & $68 \pm 34$ \\
\midrule
Connected components, accepted walls & $1.0 \pm 0.0$ & $1.0 \pm 0.0$ \\
Non-manifold edges, accepted walls   & $0.0 \pm 0.0$ & $0.0 \pm 0.0$ \\
Boundary-loop agreement              & 14/14 & 23/23 \\
\midrule
Matched ports                        & 1176 & 998 \\
Median port-centre error (mm)        & 0.120 & 0.127 \\
Ports exceeding tolerance            & 0/1176 & 0/998 \\
\midrule
Self-intersecting pairs, median      & 6 & 4 \\
\midrule
Reached generation-8 cap             & 14/15 & 20/27 \\
Grid spacing $h$ (mm)                & 0.45--0.70 & 0.40--0.60 \\
\bottomrule
\end{tabular}
\end{table}

\subsection{Baseline comparison}
\label{sec:res-baselines}

\begin{figure*}[!t]
\centering
\includegraphics[width=0.88\textwidth]{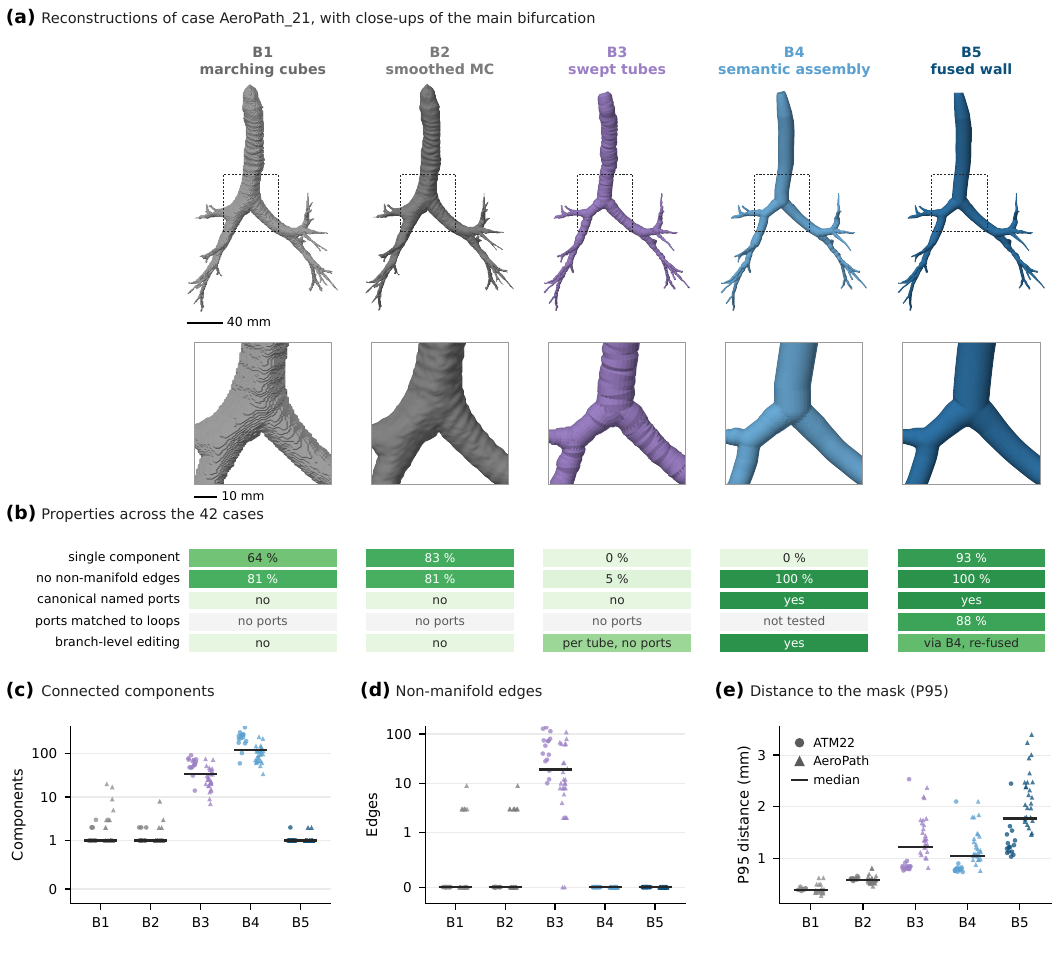}
\caption{Baseline comparison. (a)~The five strategies applied to case AeroPath\_21, with close-ups of the main
bifurcation (dashed box): marching cubes on the mask (B1) reproduces the voxel steps and tracheal rings, smoothing (B2)
softens them, independent swept tubes (B3) overlap at every junction, the semantic assembly (B4) consists of
overlapping labelled primitives, and the fused wall (B5) is a single surface. (b)~Properties across the 42 cases:
percentages are the fraction of cases in which a property holds; named ports exist only in B4 and B5, port-to-loop
matching is tested only on B5, and B5 is edited by re-fusing an edited B4 assembly. (c)~Connected components,
(d)~non-manifold edges and (e)~95th-percentile distance from the surface to the mask, per case; bars mark the medians.}
\label{fig:baselines}
\end{figure*}

The five strategies of Section~\ref{sec:evaluation} were built and evaluated on
the same 42 cases (Table~\ref{tab:baselines}, Fig.~\ref{fig:baselines}). All five produced a surface in every case.
Rejected B5 cases are retained in the topology and distance statistics; the five B5 rejections were the same cases as
in Section~\ref{sec:res-acceptance}, reproduced through a separate evaluation script.

Marching cubes on the mask, with or without smoothing (B1, B2), follows the voxel boundary closely but defines no
named ports, so port-to-loop matching is not applicable; the few boundary loops it produces are incidental openings.
On AeroPath it was not always a single manifold component, and smoothing reduced the component count without removing
non-manifold edges.

The swept tubes (B3) use the same centreline--radius scaffold without shared ports or implicit fusion. The capped
per-branch tubes overlap at junctions, produce no open boundaries and average 39.6 components and 36.9 non-manifold
edges per case, so the scaffold alone does not yield a single surface.

The semantic assembly (B4) had zero non-manifold edges in every case but is multi-component by design, being a
collection of overlapping labelled primitives; its ports are stored but not tested against external loops. Among the
five strategies, only B5 combines named ports with a single-component, edge-manifold wall whose ports can be checked
against boundary loops, and it met all three conditions in 37 of 42 cases. The differences are categorical, so they
are not combined into a composite score.

B5 had the largest surface-to-mask distance of the five strategies (pooled P95 1.77\,mm) and the highest
inside-mask fraction. B1 is the isosurface of the mask itself and is therefore a lower bound on this distance, whereas
B5 is regenerated from smoothed centrelines with a radius floor, which moves it outward at thin branches and inward
where smoothing shortens a curve. Cross-sectional calibre is assessed in Section~\ref{sec:res-fidelity}.

\begin{table*}[!t]
\centering
\caption{Baseline comparison of five reconstruction strategies on the same 42 cases, mean\,$\pm$\,SD. Acceptance
and loop matching require named external ports and are evaluated only for B5 (n/a, not applicable). $d$ is the
unsigned distance from surface vertices to the nearest mask-boundary voxel centre. Shading marks the best value in each
cohort for components, non-manifold edges and $d$ (lower is better; ties shaded); B1 is the mask isosurface and so
bounds $d$ from below.}
\label{tab:baselines}
\small
\setlength{\tabcolsep}{3pt}
\resizebox{\textwidth}{!}{%
\begin{tabular}{llrrrrrrrr}
\toprule
Cohort & Method & Acc. & Components & Non-manifold & Loops & Loop-match &
Inside frac. & \makecell{$d$ med.\\(mm)} & \makecell{$d$ P95\\(mm)} \\
\midrule
\multirow{5}{*}{ATM22}
 & B1 raw MC        & n/a    & $1.40 \pm 0.63$   & \cellcolor{bestcell}$0.00 \pm 0.00$ & $0.20 \pm 0.56$   & n/a   & 0.650 & 0.391 & \cellcolor{bestcell}0.391  \\
 & B2 smoothed MC   & n/a    & $1.20 \pm 0.41$   & \cellcolor{bestcell}$0.00 \pm 0.00$ & $0.20 \pm 0.56$   & n/a   & 0.539 & \cellcolor{bestcell}0.390 & 0.594 \\
 & B3 swept tube    & n/a    & $57.80 \pm 19.20$ & $60.67 \pm 41.03$& $0.00 \pm 0.00$   & n/a   & 0.671 & 0.441 & 0.842 \\
 & B4 semantic      & n/a    & $225.67 \pm 81.43$& \cellcolor{bestcell}$0.00 \pm 0.00$ & $302.93 \pm 108.83$& n/a  & 0.705 & 0.429 & 0.793 \\
 & \textbf{B5 fused}& 14/15  & \cellcolor{bestcell}$1.07 \pm 0.26$ & \cellcolor{bestcell}$0.00 \pm 0.00$ & $78.47 \pm 27.43$ & 14/15 & 0.781 & 0.451 & 1.220 \\
\midrule
\multirow{5}{*}{AeroPath}
 & B1 raw MC        & n/a    & $3.11 \pm 4.78$   & $1.11 \pm 2.06$  & $0.30 \pm 0.47$   & n/a   & 0.477 & \cellcolor{bestcell}0.352 & \cellcolor{bestcell}0.371  \\
 & B2 smoothed MC   & n/a    & $1.41 \pm 1.39$   & $1.11 \pm 2.06$  & $0.30 \pm 0.47$   & n/a   & 0.532 & 0.370 & 0.573 \\
 & B3 swept tube    & n/a    & $29.56 \pm 17.11$ & $23.74 \pm 29.67$& $0.00 \pm 0.00$   & n/a   & 0.745 & 0.441 & 1.348 \\
 & B4 semantic      & n/a    & $103.96 \pm 47.73$& \cellcolor{bestcell}$0.00 \pm 0.00$ & $140.00 \pm 64.28$& n/a   & 0.759 & 0.433 & 1.127 \\
 & \textbf{B5 fused}& 23/27  & \cellcolor{bestcell}$1.07 \pm 0.27$ & \cellcolor{bestcell}$0.00 \pm 0.00$ & $37.11 \pm 16.55$ & 23/27 & 0.846 & 0.553 & 2.067 \\
\midrule
\multirow{5}{*}{Pooled}
 & B1 raw MC        & n/a    & $2.50 \pm 3.92$   & $0.71 \pm 1.73$  & $0.26 \pm 0.50$   & n/a   & 0.493 & \cellcolor{bestcell}0.376 & \cellcolor{bestcell}0.386  \\
 & B2 smoothed MC   & n/a    & $1.33 \pm 1.14$   & $0.71 \pm 1.73$  & $0.26 \pm 0.50$   & n/a   & 0.534 & 0.382 & 0.585 \\
 & B3 swept tube    & n/a    & $39.64 \pm 22.34$ & $36.93 \pm 38.13$& $0.00 \pm 0.00$   & n/a   & 0.707 & 0.441 & 1.225 \\
 & B4 semantic      & n/a    & $147.43 \pm 84.81$& \cellcolor{bestcell}$0.00 \pm 0.00$ & $198.19 \pm 113.61$& n/a  & 0.732 & 0.432 & 1.047 \\
 & \textbf{B5 fused}& 37/42  & \cellcolor{bestcell}$1.07 \pm 0.26$ & \cellcolor{bestcell}$0.00 \pm 0.00$ & $51.88 \pm 28.86$ & 37/42 & 0.820 & 0.501 & 1.772 \\
\bottomrule
\end{tabular}%
}
\end{table*}

\subsection{Geometric fidelity}
\label{sec:res-fidelity}

\paragraph{Lumen calibre against the segmentation}
Stations were placed every 1.0\,mm of arc length along each branch centreline
in all 42 cases, excluding stations within $2r$ of a junction node where a
plane orthogonal to one centreline is ill-defined. Of 89\,096 stations placed,
63\,159 yielded a measurement on both the wall and the mask.

\begin{figure*}[!t]
\centering
\includegraphics[width=\textwidth]{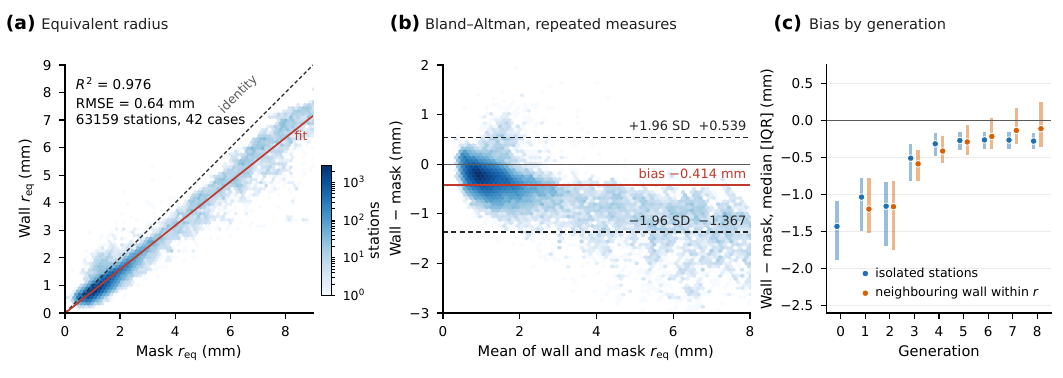}
\caption{Lumen calibre of the fused wall against the segmentation, at 63\,159 stations in 42 cases. (a)~Equivalent
radius of the wall against that of the mask, with the identity (dashed) and the least-squares fit (red); colour is
station density. (b)~Bland--Altman plot with the mean bias and the 95\,\% limits of agreement for repeated
measurements per case. (c)~Median and interquartile range of the difference by generation, for isolated stations and
for the 3.0\,\% of stations where another branch's floored wall lies within the station's own radius.}
\label{fig:fidelity}
\end{figure*}

Wall and mask equivalent radii were strongly correlated ($R^2=0.976$; Fig.~\ref{fig:fidelity}a) but showed a
systematic bias: the reconstructed lumen was narrower by a mean of 0.414\,mm (RMSE 0.638\,mm; 95\,\% limits of
agreement $-1.367$ to $+0.539$\,mm; Fig.~\ref{fig:fidelity}b). Per case, the median $R^2$ was 0.981 and the median
bias $-0.409$\,mm.

The bias follows from the representation. The tapered primitives are circular and take their radius from the
distance transform, which gives the inscribed radius, whereas the segmented lumen is not circular: the ratio of
maximum to minimum in-plane ray length on the mask averaged 1.9. A circular wall at the inscribed radius therefore
under-represents the equivalent radius of a non-circular section. The fused wall does not recover lumen eccentricity,
and the narrowing should be taken into account when the geometry is used for flow simulation.

Where two branches run close together, the hard union can let a neighbouring branch's primitive set the wall
(Section~\ref{sec:fused}). At the 3.0\,\% of stations (1873) where another branch's floored wall lay within the
station's own radius, agreement was looser ($R^2=0.886$; limits of agreement $-1.651$ to $+0.878$\,mm against
$-1.357$ to $+0.527$\,mm elsewhere; Fig.~\ref{fig:fidelity}c), while the mean bias was similar ($-0.386$ against
$-0.415$\,mm).

\paragraph{Branching statistics}
Branching statistics were computed on the 36 accepted cases with at least four
Strahler orders, covering 4502 branches. Branch length is the chord between successive branch points, diameter the
mean over the central half of the branch, and branches separated by a bifurcation that lost a child to spur pruning
are kept distinct (Section~\ref{sec:evaluation}). Table~\ref{tab:morphometry} reports means over cases;
Table~\ref{tab:ld-generation} pools branches by generation.

Branch angle and daughter length ratio were nearly identical in the two cohorts, whereas the diameter ratios and the
Strahler diameter ratio differed more (Table~\ref{tab:morphometry}; Supplementary Fig.~\ref{fig:s-branching}).
Comparison with cast-derived reference morphometry~\cite{weibel1963morphometry,horsfield1971models} is not attempted
here, because the generation cap, the pruning thresholds and the CT resolution truncate the tree in ways those data do
not share.

The length-to-diameter ratio depends strongly on generation. Its case mean was 6.07
(Table~\ref{tab:morphometry}); pooled by generation (Table~\ref{tab:ld-generation}), it was 2.83--3.38 for
generations 1--3 and rose monotonically from generation~3 to 7.73 at generation~8.

Because removing terminal branches shorter than 8.0\,mm selects on length, its effect on the distal rise was
tested by relaxing the length threshold fourfold on a twelve-case subset. Generation-8 $L/D$ fell by 4.2\,\%, less in
relative terms than at generations 2, 3 and 6, and the rise remained (Supplementary Fig.~\ref{fig:s-branching}f). The
removed branches had a mean length of 4.8--5.4\,mm at every generation.

\begin{table}[t]
\centering
\footnotesize
\setlength{\tabcolsep}{3pt}
\caption{Branching statistics of the reconstructed trees, mean\,$\pm$\,SD across cases. Length is the chord between
successive branch points; ordering ratios exclude the highest Strahler order.}
\label{tab:morphometry}
\begin{tabular}{lrrr}
\toprule
Measure & ATM22 ($n{=}14$) & AeroPath ($n{=}22$) & Pooled ($n{=}36$) \\
\midrule
\multicolumn{4}{l}{\emph{Per-bifurcation statistics}} \\
Branch angle $\theta$ ($^{\circ}$)        & $42.2 \pm 3.8$ & $42.1 \pm 2.6$ & $42.1 \pm 3.1$ \\
Length-to-diameter $L/D$                   & $5.91 \pm 1.02$ & $6.17 \pm 1.06$ & $6.07 \pm 1.04$ \\
$D_{\mathrm{child}}/D_{\mathrm{parent}}$   & $0.790 \pm 0.019$ & $0.721 \pm 0.056$ & $0.748 \pm 0.057$ \\
$D_{\mathrm{minor}}/D_{\mathrm{major}}$    & $0.826 \pm 0.025$ & $0.757 \pm 0.046$ & $0.784 \pm 0.052$ \\
$L_1/L_2$                                  & $0.613 \pm 0.031$ & $0.609 \pm 0.050$ & $0.610 \pm 0.043$ \\
\midrule
\multicolumn{4}{l}{\emph{Ordering ratios}} \\
Strahler branching $R_{bS}$               & $2.566 \pm 0.157$ & $2.769 \pm 0.337$ & $2.690 \pm 0.296$ \\
Strahler length $R_{lS}$                  & $1.185 \pm 0.112$ & $1.327 \pm 0.208$ & $1.272 \pm 0.188$ \\
Strahler diameter $R_{dS}$                & $1.473 \pm 0.057$ & $1.687 \pm 0.117$ & $1.603 \pm 0.144$ \\
Horsfield branching $R_{bH}$              & $1.522 \pm 0.089$ & $1.485 \pm 0.102$ & $1.499 \pm 0.098$ \\
\midrule
Strahler orders                            & $5.71 \pm 0.47$ & $4.68 \pm 0.57$ & $5.08 \pm 0.73$ \\
Branches per case                          & $186 \pm 70$ & $86 \pm 40$ & $125 \pm 72$ \\
\bottomrule
\end{tabular}
\end{table}

\begin{table}[t]
\centering
\caption{Length-to-diameter ratio by anatomical generation, pooled over branches of the 36 cases in
Table~\ref{tab:morphometry}.}
\label{tab:ld-generation}
\begin{tabular}{lrrrr}
\toprule
Generation & ATM22 & AeroPath & Pooled & $n$ (pooled) \\
\midrule
1 & 3.34 & 3.41 & 3.38 & 83 \\
2 & 2.69 & 2.93 & 2.83 & 146 \\
3 & 2.71 & 3.70 & 3.31 & 291 \\
4 & 3.57 & 5.12 & 4.45 & 595 \\
5 & 5.54 & 6.78 & 6.10 & 958 \\
6 & 6.72 & 7.88 & 7.15 & 1025 \\
7 & 7.17 & 7.84 & 7.40 & 837 \\
8 & 7.44 & 8.67 & 7.73 & 529 \\
\bottomrule
\end{tabular}
\end{table}

\subsection{Ablation and computational cost}
\label{sec:res-ablation}

\begin{figure*}[!t]
\centering
\includegraphics[width=\textwidth]{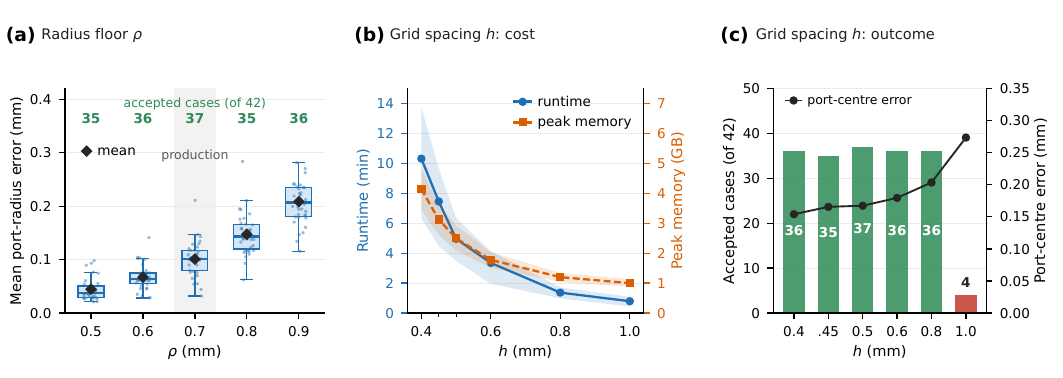}
\caption{Sensitivity to the radius floor and the grid spacing, each swept over all 42 cases with every other setting
at its production value (462 reconstructions). (a)~Mean port-radius error per accepted case against the radius floor
$\rho$; boxes show the interquartile range, diamonds the mean, and the numbers the accepted cases. (b)~Runtime and peak
memory against the grid spacing $h$ (mean over cases; shading, interquartile range). (c)~Accepted cases (bars) and
mean port-centre error of accepted cases (line) against $h$; at $h=1.00$\,mm (red) the grid no longer resolves the
distal lumens.}
\label{fig:ablation}
\end{figure*}

The radius floor (five values) and the grid spacing (six values) were each swept over all 42 cases with every other
parameter at its production value, giving 462 attempted reconstructions (Fig.~\ref{fig:ablation}). Of these, 378
completed with a surface and the remainder collapsed during extraction or clean-up; acceptance was then applied to the
completed runs. Runtime and memory are averaged over all attempted runs; port errors are averaged over accepted runs
only. Ablation runs were executed as a concurrent batch, so their absolute runtimes are comparable only within the sweep;
production timing is reported separately below.

\paragraph{Radius floor}
Over $\rho\in\{0.5,0.6,0.7,0.8,0.9\}$\,mm the mean port-radius error rose monotonically from 0.044 to 0.209\,mm,
because outlets narrower than the floor are enlarged (Supplementary Fig.~\ref{fig:s-port-radius}). The other measures
changed little: port-centre error by less than 0.01\,mm, mean runtime by less than 5\,\% and peak memory by less than
1\,\%. Between 36 and 38 runs completed and 35 to 37 were accepted at each value. The production value of 0.7\,mm,
fixed before the evaluation (Section~\ref{sec:data}), gave the largest count, but a spread of two cases does not
identify an optimum; the floor trades realised outlet radius for meshable sub-millimetre outlets, a cost the
acceptance criteria do not detect.

\paragraph{Grid spacing}
Over $h\in\{0.40,0.45,0.50,0.60,0.80\}$\,mm, 36--38 runs completed and 35--37 were accepted at each value, the
mean port-radius error stayed between 0.097 and 0.101\,mm, and the port-centre error increased monotonically from
0.154 to 0.204\,mm. At $h=1.00$\,mm only 11 runs completed and 4 were accepted, because the grid no longer resolves
distal lumens whose diameter approaches the cell size; the port-centre error of these four was 0.274\,mm, and their
port-radius error is not comparable with the finer settings. Mean runtime fell from 619\,s at 0.40\,mm to 447, 297,
201, 82 and 47\,s at 0.45, 0.50, 0.60, 0.80 and 1.00\,mm, and mean peak memory from 4.14\,GB to 3.12, 2.50, 1.77,
1.20 and 1.00\,GB. The automatic ladder therefore selects the finest spacing within the point budget, which in
production was 0.40--0.70\,mm.

\paragraph{Computational cost}
Measured separately at the production settings (one case at a time, eight pinned threads, Intel Xeon Gold 5218R),
reconstruction from the pruned centreline graph to the validated fused wall took a median of 76.7\,s per case (IQR
67.5--91.7\,s, range 42.9--118.9\,s; median of three timed repeats after one warm-up), with a median peak resident
memory of 2.18\,GB (IQR 2.04--2.34\,GB). Voxel skeletonisation, cached per case, is excluded. Field evaluation and
surface extraction took a median of 81\,\% of the time. Runtime correlated with tree size (Pearson $r=0.74$ against
graph edges) more than with grid size ($r=0.24$), and no repeat deviated by more than 20\,\% from its case median
(Supplementary Fig.~\ref{fig:s-cost}).

\subsection{Controlled pathology generation}
\label{sec:res-pathology}

\begin{figure*}[!t]
\centering
\includegraphics[width=0.64\textwidth]{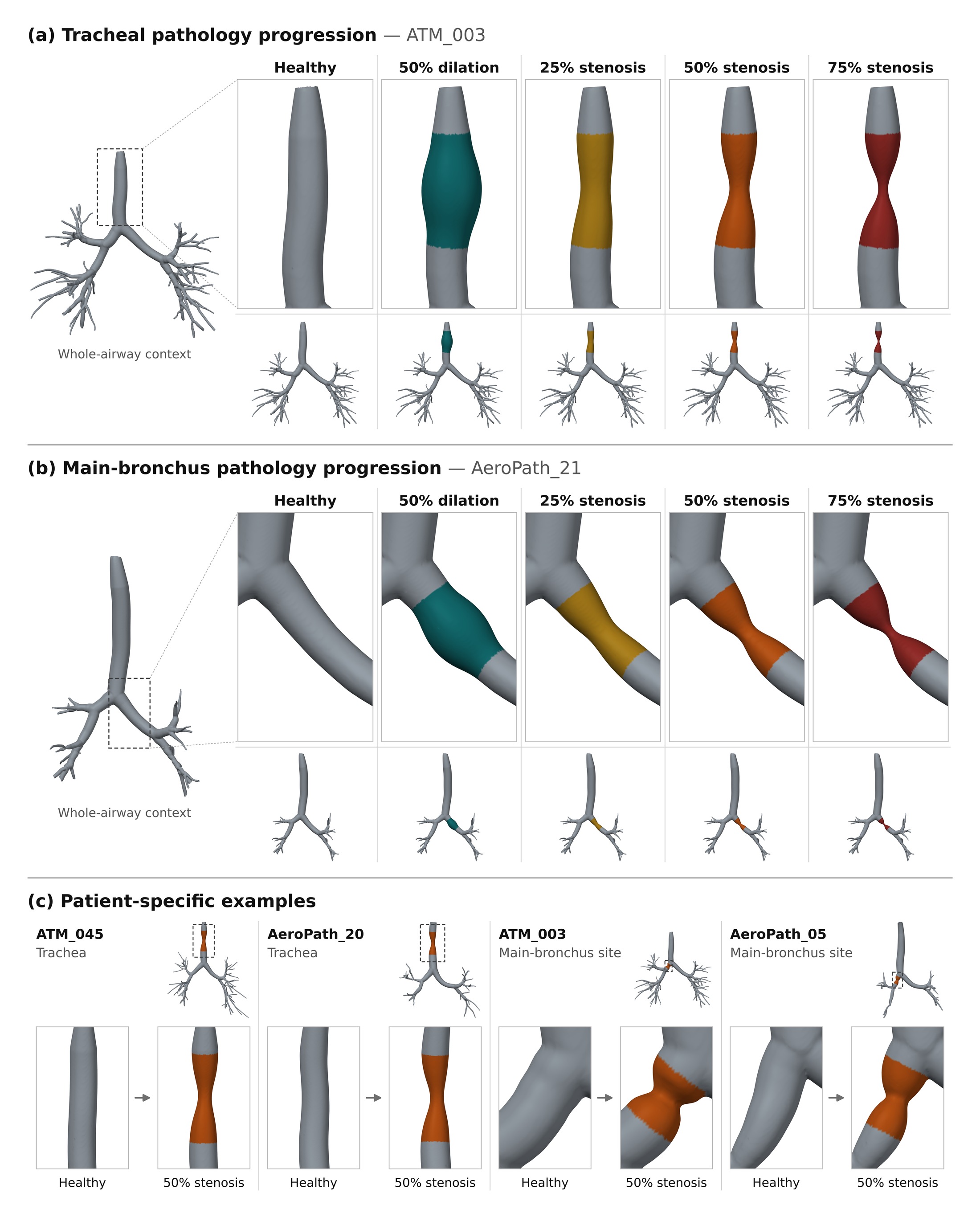}
\caption{Qualitative examples of controlled airway geometry edits. Healthy
reconstructions are compared with 50\,\% dilation and 25\,\%, 50\,\% and
75\,\% stenosis at the trachea and the algorithmically selected main-bronchus
branch. Whole-tree and close-up views highlight the edited region in colour;
the bottom row shows additional case examples. Percentages denote changes in
radius, not cross-sectional area, and do not represent a temporal progression
of disease.}
\label{fig:pathology-qualitative}
\end{figure*}

\begin{table}[t]
\centering
\caption{Controlled pathology generation across five cases and 40 variants,
measured at the commanded lesion centre on the re-fused surface. Expected area
ratio is $(q_r^{*})^2$. MAE is the mean absolute error over the ten variants in
each row.}
\label{tab:pathology}
\footnotesize
\setlength{\tabcolsep}{2.2pt}
\begin{tabular}{lrrrrrr}
\toprule
Edit & $n$ & Cmd.\ $q_r$ & Achieved $q_r$ & Radius & Exp.\ $q_A$ & CSA \\
     &     &              & median [IQR]   & MAE    &              & MAE \\
\midrule
25\,\% sten. & 10 & 0.75 & 0.7499 [0.0011] & 0.0031 & 0.5625 & 0.0047 \\
50\,\% sten. & 10 & 0.50 & 0.4992 [0.0057] & 0.0079 & 0.2500 & 0.0082 \\
75\,\% sten. & 10 & 0.25 & 0.2482 [0.0048] & 0.0108 & 0.0625 & 0.0059 \\
50\,\% dil. & 10 & 1.50 & 1.5003 [0.0009] & 0.0007 & 2.2500 & 0.0019 \\
\midrule
\multicolumn{7}{l}{\emph{By site}} \\
Trachea                & 20 & --- & --- & 0.00084 & --- & 0.00082 \\
Main bronchus          & 20 & --- & --- & 0.01034 & --- & 0.00953 \\
\bottomrule
\end{tabular}
\end{table}

Forty variants were generated from five reconstructions at two proximal sites
each --- the trachea and an algorithmically selected main-bronchus branch ---
using the lesion model of Section~\ref{sec:pathology}, alongside five healthy
baselines rebuilt under the same settings. All 45 geometries were generated and all 40 variants were analysed
(Fig.~\ref{fig:pathology-qualitative}).

\paragraph{Prescribed severity is reproduced on the fused surface}
Measured at the commanded lesion centre on the re-fused surface, the mean absolute radius-ratio error over the 40
variants was 0.0056 and the mean absolute area-ratio error 0.0052 (Table~\ref{tab:pathology},
Fig.~\ref{fig:pathology-quant}; per-variant errors in Supplementary Fig.~\ref{fig:pathology-errors-supp}a). The
worst case was the 75\,\% main-bronchus stenosis of ATM\_003, where the achieved radius ratio was 0.325 against a
commanded 0.25 (error 0.0754; area-ratio error up to 0.0572 across variants).

Tracheal edits were about an order of magnitude more accurate than main-bronchus edits (radius-ratio MAE 0.00084
against 0.01034; maximum 0.0030 against 0.0754), and the error increased with commanded stenosis severity, reaching
an MAE of 0.0108 at 75\,\% stenosis. Because the lesion width is a fixed fraction of branch length, the Gaussian is
shorter on the short main-bronchus branches and interacts more with the smoothing of Section~\ref{sec:fused}; this
association was not tested directly.

\paragraph{The perturbation stays where it was commanded}
Locality was assessed against a lesion interval fixed from the prescribed
Gaussian parameters before any reconstructed surface was examined, and
restricted to the branch shaft by a predefined margin
(Section~\ref{sec:pathology}). Outside that interval the median of the per-variant P95 deviation between edited
and healthy surfaces was 0.111\,mm (maximum P95 0.235\,mm), and the largest single deviation in the cohort was
0.300\,mm (Supplementary Fig.~\ref{fig:pathology-errors-supp}b).

\paragraph{Boundary structure is preserved}
Every edited variant met the topological acceptance criteria: a single connected component, zero non-manifold
edges and one-to-one correspondence between its external ports and boundary loops. The realised loops move slightly between independently fused surfaces:
the per-variant P95 shift was at most 0.327\,mm, and the largest shift of any
single loop was 0.411\,mm (Supplementary Fig.~\ref{fig:pathology-errors-supp}c).
In 17 of the 20 main-bronchus variants every loop was exactly unchanged, because
those edits do not alter the largest primitive radius that sets the extraction
grid. These shifts describe differences between
two independently contoured, clipped and smoothed surfaces, not movement of the
canonical ports, which the editing workflow reuses unchanged.

Boundary preservation does not imply self-intersection freedom. Every variant contained at least one
self-intersecting triangle pair, as did all five healthy baselines (2--10 pairs each;
Section~\ref{sec:res-acceptance}), so a non-zero count after editing does not by itself indicate an effect of editing.
Intersections lay on the edited branch in 10 of the 40 variants and inside the commanded lesion region in 6.

\paragraph{Scope}
The edits are geometric: the lesion is an axisymmetric modulation of a scalar radius profile and does not represent
eccentric stenosis, airway-wall mechanics, dynamic collapse, secretions or disease progression.

\begin{figure}[t]
\centering
\includegraphics[width=\columnwidth]{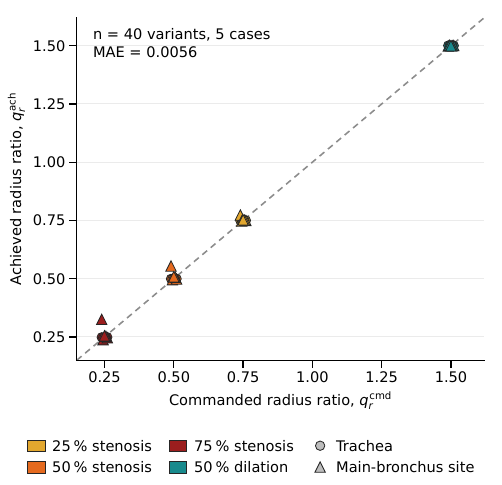}
\caption{Commanded against achieved radius ratio at the lesion centre for the
40 pathology variants (5 cases $\times$ 2 sites $\times$ 4 edits), measured on
the re-fused surface. Colour, prescribed edit; circles, trachea; triangles,
algorithmically selected main-bronchus site. The dashed line is identity.
Small horizontal offsets separate cases for display only. Quantities are defined in Eq.~\eqref{eq:s-qr}.}
\label{fig:pathology-quant}
\end{figure}

\subsection{Downstream meshing and CFD demonstration}
\label{sec:res-cfd}

\begin{table*}[!t]
\centering
\caption{Downstream CFD demonstration: mesh, set-up and convergence. Mesh metrics are from OpenFOAM's
\texttt{checkMesh} and the mesh geometry; outlet resistances follow Eqs.~\eqref{eq:outlet-resistance}
and~\eqref{eq:weibel-resistance}.}
\label{tab:cfd-setup}
\footnotesize
\begin{tabular}{@{}p{0.25\textwidth}p{0.71\textwidth}@{}}
\toprule
\multicolumn{2}{@{}l}{\emph{Geometry and mesh}}\\
Surface & capped fused wall; STL; 243\,620 triangles; mm \\
Volume mesh & ICEM CFD Tetra (Robust/Octree); 656\,650 tetrahedra, 148\,083 nodes; no prism layers \\
Global sizing & scale factor 1.3; maximum element 2.0\,mm; curvature/proximity refinement, minimum size 0.5\,mm \\
Boundaries & wall; 1 inlet (176 faces); 62 outlets (5--80 faces each, median 15.5); 3 openings $<0.7$\,mm closed as
             wall \\
Mesh quality & non-orthogonality maximum $59.9^\circ$, mean $17.3^\circ$; skewness maximum 0.65; aspect ratio maximum
               5.2 \\
Edge length & median 0.66\,mm; 5th--95th percentile 0.47--1.12\,mm; maximum 2.53\,mm \\
Transfer & boundary nodes within $4.8\times10^{-3}$\,mm of the STL; area $-1.1$\,\%; volume $-0.85$\,\% \\
\midrule
\multicolumn{2}{@{}l}{\emph{Physics and numerics (OpenFOAM v1912, \texttt{simpleFoam})}}\\
Fluid & air, incompressible; $\rho_f=1.204$\,kg\,m$^{-3}$; $\mu=1.825\times10^{-5}$\,Pa\,s; steady, laminar \\
Inlet & 0.1\,L\,s$^{-1}$; smooth profile, peak/mean 1.75; $U=1.67$\,m\,s$^{-1}$; $D_h=8.71$\,mm; $Re\approx960$ \\
Outlets & $p_k=R_kQ_k$; Weibel subtree from generation $g+1$ to 16; $g=5$--12;
          $R_k=2.3\times10^{5}$--$4.8\times10^{6}$\,Pa\,s\,m$^{-3}$ \\
Wall & stationary, no-slip \\
Solver & SIMPLEC; bounded \texttt{linearUpwind} convection; outlet pressures updated every 20 iterations \\
Convergence criterion & boundary-condition error $<1$\,\% and outlet-share change $<0.01$ pt over two updates \\
\midrule
\multicolumn{2}{@{}l}{\emph{Result}}\\
Convergence & 380 iterations (17 updates); boundary-condition error 0.19\,\%; share change 0.005 pt;
              residuals $\le4.4\times10^{-5}$; outflow/inflow $1-10^{-6}$; no outlet backflow \\
Pressure & inlet 3.37\,Pa; outlets 0.96--2.49\,Pa; maximum 6.0\,Pa at the carina \\
Flow split & $-x$ lung 60.2\,\% (34 outlets), $+x$ lung 39.8\,\% (28 outlets); outlets 0.30--7.6\,\% \\
\bottomrule
\end{tabular}
\end{table*}

\begin{figure*}[!t]
\centering
\includegraphics[width=0.78\textwidth]{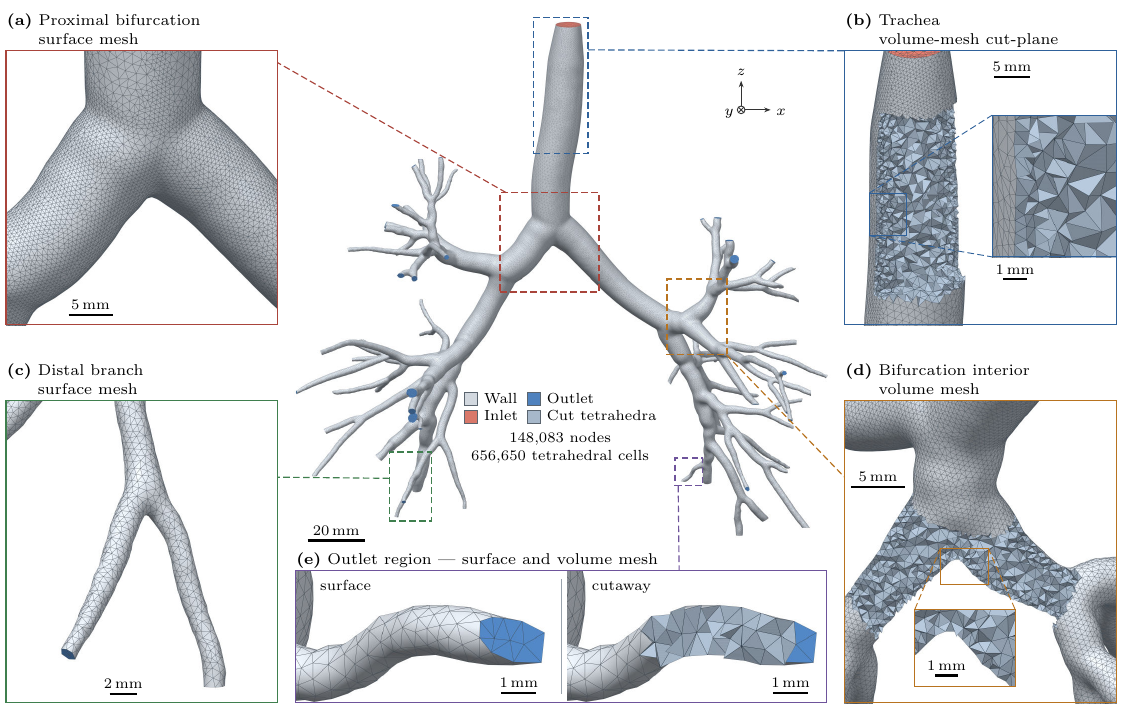}
\caption{Volume mesh of the CFD demonstration case (ICEM CFD Robust/Octree; 656\,650 tetrahedra, no prism layers).
Enlarged views: (a)~main bifurcation, (b)~tracheal cut plane, (c)~distal branch, (d)~bifurcation interior and
(e)~outlet region.}
\label{fig:mesh}
\end{figure*}

\begin{figure*}[!t]
\centering
\includegraphics[width=\textwidth]{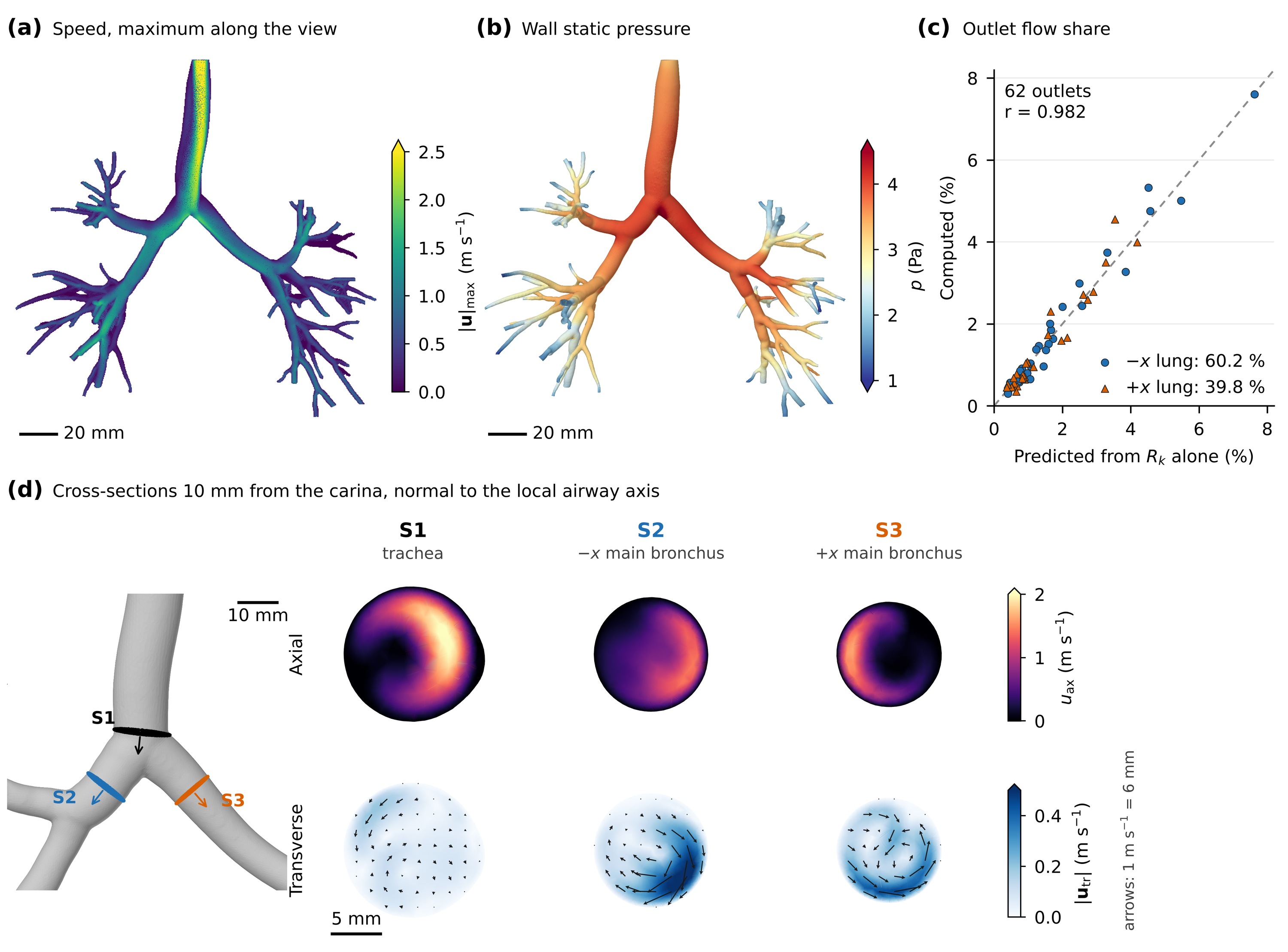}
\caption{OpenFOAM solution at the stopping criterion: steady laminar flow at 0.1\,L\,s$^{-1}$ with a resistance boundary condition on
each of the 62 outlets. (a)~Largest speed along the viewing direction (Eq.~\eqref{eq:s-mip}); (b)~wall static pressure,
relative to the distal end of the modelled subtrees; (c)~computed flow share of each outlet against the share predicted
from its resistance alone (Eq.~\eqref{eq:s-share}); (d)~sections S1--S3, 10\,mm from the carina and normal to the local
airway axis (locator, left), viewed along the flow: axial velocity (top) and transverse velocity, magnitude and in-plane
vectors (bottom; Eq.~\eqref{eq:s-decomp}). $\pm x$ are CT image axes.}
\label{fig:cfd}
\end{figure*}

\paragraph{Volume mesh and transfer}
The volume mesh (Fig.~\ref{fig:mesh}, Table~\ref{tab:cfd-setup}) contains 656\,650 tetrahedra and 148\,083 nodes, and
\texttt{checkMesh} reported no errors: maximum non-orthogonality $59.9^\circ$ (mean $17.3^\circ$), maximum skewness
0.65 and maximum aspect ratio 5.2. Edge lengths have a median of 0.66\,mm, span 0.47--1.12\,mm between the 5th and
95th percentiles, and reach a maximum of 2.53\,mm.

All 70\,665 boundary nodes lie on the exported STL surface (median distance $1.1\times10^{-5}$\,mm, maximum
$4.8\times10^{-3}$\,mm). The boundary area is 1.1\,\% smaller than that of the STL and the enclosed volume 0.85\,\%
smaller, consistent with coarser facets on a curved wall. The outlets are the most coarsely resolved boundaries: the
62 open outlets carry between 5 and 80 faces each (median 15.5), and 25 of them have ten faces or fewer.

\paragraph{Convergence}
The stopping criteria of Section~\ref{sec:cfd} were met after 380 solver iterations (17 outlet updates;
Supplementary Fig.~\ref{fig:cfd-conv}): at the final update the largest outlet boundary-condition error was 0.19\,\%
and the largest change in any outlet flow share 0.005 percentage points. The scaled initial residuals had fallen by
more than two orders of magnitude, to $1.7\times10^{-5}$--$4.4\times10^{-5}$ for velocity and $4.3\times10^{-5}$ for
pressure, and were still decreasing when the run stopped. Total outflow matched inflow to within a relative
$10^{-6}$, and no outlet had reverse flow.

\paragraph{Flow field}
The inflow enters the trachea as a jet (Fig.~\ref{fig:cfd}a) with a peak speed of 2.98\,m\,s$^{-1}$. The
jet runs along one side of the widening trachea and its core reaches the carina, where it stagnates and produces the
highest pressure in the domain, 6.0\,Pa, at the carinal ridge (Fig.~\ref{fig:cfd}b). Downstream of the carina the flow divides
into the main bronchi, and speeds fall with each generation; 1.2\,\% of the projected airway area in Fig.~\ref{fig:cfd}a exceeds
the 2.5\,m\,s$^{-1}$ colour limit.

The cross-sections show how the jet is redistributed at the bifurcation (Fig.~\ref{fig:cfd}d). At S1, 10\,mm
above the carina, the axial velocity peaks at 2.1\,m\,s$^{-1}$, 2.9 times its mean of 0.73\,m\,s$^{-1}$, and the
fastest tenth of the section lies 3.1\,mm from its centroid, towards the carina; 6.8\,\% of the section carries reverse
axial flow. Transverse velocities there are small, 9.5\,\% of the mean axial velocity. In both main bronchi the axial
maximum is displaced towards the carinal (inner) wall, by 3.5\,mm in both S2 and S3, about 0.65 of the equivalent
radius, and secondary flow is stronger: the mean transverse velocity is 30\,\% of the mean axial velocity in S2 and
38\,\% in S3, which also has reverse flow over 6.5\,\% of its area. Along each main bronchus the secondary flow decays
with distance from the carina (Supplementary Fig.~\ref{fig:cfd-profiles}).

The mean pressure needed at the inlet was 3.37\,Pa. Outlet pressures ranged from 0.96 to 2.49\,Pa, and their
flow-weighted mean was 1.83\,Pa, so about 1.5\,Pa of the drop occurs within the reconstructed airways and the
remainder in the modelled distal subtrees.

\paragraph{Flow partition}
The $-x$ lung (right in the LPS frame), drained by 34 outlets, received 60.2\,\% of the inflow and the $+x$ lung, with
28 outlets, 39.8\,\%. Individual outlets carried between 0.30 and 7.6\,\% of the inflow, and 32 of the 62 carried less
than 1\,\% each. The computed shares followed those predicted from the outlet resistances alone closely
(Fig.~\ref{fig:cfd}c; $r=0.982$), deviating by at most 1.0 percentage point, which reflects the resistance of the
resolved airways. Per-outlet results are given in Supplementary Table~\ref{tab:cfd-outlets}.

\paragraph{Scope of the demonstration}
One accepted fused wall was capped, meshed and solved with a separate resistance condition at each of its 62 modelled
outlets, and the coupled problem met its stopping criteria with mass conservation and no outlet backflow. This shows
that the reconstructed geometry can enter a standard meshing and solver chain. It does not establish that every
accepted wall can be meshed without repair, and, being a single steady laminar case at one low flow rate on one mesh
without a mesh-independence study, it is not a validation of airflow predictions.

\section{Discussion}
\label{sec:discussion}

\subsection{Semantic consistency and fused-surface validity}
\label{sec:disc-guarantee}

The shared-port representation makes interface consistency a structural property. Adjacent blocks reference a
single port object, so their centres, normals, radii and frames agree without fitting, snapping or repair, and every
inlet and outlet is a named object with a position, orientation and radius in physical coordinates. This guarantees the
internal consistency of the semantic model, but it does not guarantee anything about the surface subsequently
generated from it. The two properties must therefore be assessed separately. Port coincidence in the assembly holds by
construction. Port-to-loop correspondence on the fused wall is an empirical property: it held in 37 of 42 cases and
failed in five. Because this check can fail, it shows where the construction breaks down instead of assuming that it
does not.

Where correspondence held, the exported port manifest describes the realised boundary loops rather than only the
intended ones, with sub-millimetre agreement for almost all ports. This is the main practical value of the
representation for downstream modelling: boundary conditions can be associated with named ports and traced back to
branches of the editable model. The residual port-to-loop displacement did not decrease when the field was
re-evaluated at one third of the production grid spacing. It is therefore better interpreted as a difference between
two constructions of the same interface, the analytic port plane and the emergent zero level set, than as a sampling
error; this is why the acceptance tolerance includes an absolute term.

\subsection{Geometric approximations of implicit-union reconstruction}
\label{sec:disc-envelope}

A hard minimum over tapered primitives was used instead of smooth blending, which can enlarge the lumen near
junctions. Because the minimum is taken over all nearby primitives, however, the wall follows whichever primitive is
closest, including one belonging to an anatomically unrelated branch. Where two branches run close together, one
lumen can be set by the envelope of the other, and at ports two outlets can merge into one boundary loop, as in one of
the five rejected cases. Elsewhere the effect was local: agreement with the mask was looser at the small fraction of
stations near another branch, while the mean bias was similar. The behaviour was not reduced by grid refinement and
was not removed by restricting the primitives a port may consider. Resolving it would require a branch-aware field in
which each point is assigned to an owning branch, which brings its own validation requirements at junctions. Because
unions of centreline-based primitives are also used in vascular modelling~\cite{antiga2008framework}, the same
limitation may apply there.

The second approximation is systematic. The reconstructed lumen was narrower than the segmentation by a mean of
0.414\,mm, consistent with circular primitives sized by the inscribed distance-transform radius in lumens that are not
circular. A narrower lumen increases flow resistance, so simulations on this geometry would be biased towards higher
resistance unless the narrowing is corrected or accounted for. The circular cross-section is a deliberate choice: it
gives every interface a single radius and frame and supports the parametric editing evaluated here. Elliptical
primitives or swept cross-sections recovered from the mask could represent eccentricity, but they would require a
different interface definition. The radius floor adds a further, localised departure from the image: outlets narrower
than 0.7\,mm are enlarged in the fused wall, although the semantic model retains the unfloored radii.

Branching statistics were sensitive to measurement conventions. Defining length as arc length rather than chord
length, or allowing pruning to merge consecutive branches, changed the length-to-diameter ratio substantially
(\hyperref[sec:supp-note]{Supplementary Note~S1}). Even with consistent definitions, the ratio rose with generation, and this
could not be attributed to the terminal-pruning threshold. Whether the rise reflects anatomy, limited distal
resolution or the diameter convention cannot be determined from these data. Resolving it would require
generation-resolved reference measurements made with the same conventions, for example from automated bronchial
parameter extraction~\cite{dudurych2023reproducibility} or topological airway
descriptors~\cite{kaji2024persistent}.

\subsection{Relation to existing approaches}
\label{sec:disc-prior}

Pipelines that extract and repair an isosurface produce simulation domains directly from segmentations, including
automated airway meshing~\cite{lauria2022airwaymesh} and subject-specific CT-to-CFD workflows with automated inlet
and outlet processing~\cite{bartol2025subjectspecific}. In these approaches, as in baselines B1 and B2, boundaries are
identified on the processed surface rather than defined in a branch-level model. Centreline-and-diameter
reconstruction~\cite{tawhai2004ct,miyawaki2017automatic} recovers a branching scaffold from CT; baseline B3 shows that
such a scaffold, swept into independent tubes, does not form a single surface without an explicit joining step.
Vascular toolkits combine centreline analysis, surface editing and meshing~\cite{antiga2008framework,marchandise2013mesh},
and procedural generators construct airways from rules or specifications rather than from an individual image,
including volume-filling growth~\cite{tawhai2000generation}, space-filling tree models~\cite{kitaoka1999three},
AVATREE~\cite{nousias2020avatree}, RespGeomLib~\cite{wasalathilaka2026respgeomlib} and hybrid CT-informed stochastic
growth~\cite{spasov2026bronchialtrees}. Improved segmentation~\cite{zhang2023multisite,wang2024airwaytopology} provides
better input masks but not an editable, port-labelled geometry. Studies of stenosis effects on airflow typically
construct each modified geometry individually~\cite{ding2023stenotic,michauddorko2024stenosis}; here, edits are
applied locally to the radius profile and their realised severity and extent are measured on the regenerated
surface.

The individual elements of the framework are not new. Its contribution lies in combining, for CT-derived airways, an
editable branch-level model in physical coordinates, a fused wall with one named boundary per port, and an explicit
test that each port is realised as its own boundary loop, evaluated automatically on two public cohorts.

\subsection{Limitations and future directions}
\label{sec:disc-limitations}

Topological acceptance does not ensure a surface free of self-intersections: intersecting triangle pairs were present
in 40 of 42 fused walls and in every pathology variant and healthy baseline. An accepted wall may therefore still
require repair before volume meshing, and direct meshability should not be inferred from acceptance. A repair stage
validated against self-intersection freedom, followed by meshing tests across the cohort, is needed before stronger
claims about mesh readiness can be made.

Five cases were rejected. One is explained by the envelope behaviour described above; the other four contain surplus
boundary loops introduced at the clipping stage, for which no common mechanism was identified. During development,
the defects that were found and corrected all lay in surface clean-up rather than in the geometric core
(\hyperref[sec:supp-note]{Supplementary Note~S1}), suggesting that validation after each clean-up step, together with a
branch-aware union, is the most direct route to fewer failures.

The evaluation used 42 reference annotations rather than automatic segmentations, with 15 of several hundred ATM22
cases, and no oblique acquisitions or slices thicker than 1.25\,mm; performance on noisier masks and on clinical
thick-slice CT remains untested. Reconstruction is limited to eight generations and prunes short or thin terminal
branches, so the result is a conducting-airway model. Such trees could serve as patient-specific proximal scaffolds for
generative models that extend the airway beyond CT resolution~\cite{tawhai2000generation,spasov2026bronchialtrees}.

The CFD demonstration used one case, one mesh and a single steady laminar flow of 0.1\,L\,s$^{-1}$, without a
mesh-independence study or near-wall prism layers. The outlet resistances were derived from a symmetric Weibel subtree
scaled to each outlet diameter, so a one-generation change alters a resistance by about 15--17\,\%. Outlet patches
were defined from connected openings rather than mapped to the named ports of the manifest. The demonstration
therefore shows that a reconstructed wall can enter a standard solver chain, but it does not validate the predicted
flow. Validation would require comparison with measurements, as in image-based deposition
studies~\cite{longest2016validating,sadafi2024validation}. Port-based patch assignment, transient and higher-flow
simulations, and mesh-sensitivity analysis are natural next steps.

\section{Conclusion}
\label{sec:conclusion}

\RespGeomLib{} converts CT airway masks into two linked outputs: an editable assembly of labelled pipe and junction
blocks that share canonical ports, and a fused lumen surface whose acceptance requires each port to be matched to its
own boundary loop. On 42 reference masks from two cohorts, 37 reconstructions met this criterion, with sub-millimetre
port-to-loop agreement throughout, and the five failures were traced to identifiable surface-level mechanisms rather
than to port placement. The fused wall follows the segmented lumen closely in shape but represents it as a circular
tube at the inscribed radius, giving a systematic narrowing of about 0.4\,mm that users should account for. The same
representation supports localised, quantitatively reproduced radius edits and provides one named boundary per outlet
for downstream simulation. Self-intersection repair, outlet-to-port mapping in the solver and validation of flow
predictions are outside the present scope and are the next steps towards a complete image-to-simulation workflow.

\section*{Code and data availability}

The source code, environment specification and per-case configuration files will be released in a public repository upon publication. Dataset availability follows the access conditions of the original ATM22 and AeroPath data providers.

\section*{Declaration of competing interest}

The authors declare that they have no known competing financial interests or personal relationships that could have appeared to influence the work reported in this paper.

\balance
\bibliographystyle{elsarticle-num}
\bibliography{references}

\clearpage
\onecolumn

\section*{Supplementary Methods}
\label{sec:supplementary-methods}
\setcounter{figure}{0}
\renewcommand{\thefigure}{S\arabic{figure}}
\setcounter{table}{0}
\renewcommand{\thetable}{S\arabic{table}}

These Supplementary Methods give the case-level data, the reasoning behind design choices and the complete
parameter list that the main Methods summarise. SM1--SM9 follow the order of the main Methods; SM10 defines the
quantities shown in the figures.

\begin{center}
\captionof{table}{Complete list of reconstruction parameters, their production values and purpose. All values were fixed
before the evaluation run. $h$, grid spacing; tol, port tolerance; $r$, port or node radius;
$\ell_{\mathrm{seg}}^{\max}$, longest segment.}
\label{tab:params-full}
\small
\begin{tabular}{@{}lll@{}}
\toprule
Parameter & Value & Purpose \\
\midrule
\multicolumn{3}{@{}l}{\emph{Graph extraction}} \\
Largest-component connectivity & 18 & remove isolated mask fragments \\
Skeleton / graph connectivity & 26 & thinning and chain tracing \\
Bounding-box margin & 8 voxels & crop with context \\
Minimum chain length & 3 voxels & discard skeleton noise \\
\midrule
\multicolumn{3}{@{}l}{\emph{Pruning and extent}} \\
Spur length threshold & 8.0\,mm & prune terminal spurs \\
Spur radius threshold & 0.75\,mm & prune terminal spurs \\
Maximum generation depth & 8 & standardise airway extent \\
\midrule
\multicolumn{3}{@{}l}{\emph{Blocks}} \\
Cut factor $\alpha$ & 2.0 (cut $\le0.4\,\ell_e$) & junction stand-off distance \\
Curved-pipe criteria & arc/chord $>1.03$ or deviation $>5$\,\% & curved/straight classification \\
Curved-pipe circumferential samples & 96 & semantic tube resolution \\
\midrule
\multicolumn{3}{@{}l}{\emph{Fused field}} \\
Smoothing widths & $0.08\,\ell_e$ (centreline), $0.18\,\ell_e$ (radius) & profile smoothing \\
Endpoint blend & terminal 20\,\% of each edge & restore port values \\
Radius floor $\rho$ & 0.7\,mm & distal meshability (fused wall only) \\
Union breadth $K$ & 24 & nearest segments in the field \\
Narrow-band half-width & $r_{\max}+\tfrac12\ell^{\max}_{\mathrm{seg}}+3h$ & bound field evaluation \\
Grid margin & $\max(2r_{\max}+3h,\,8\,\mathrm{mm})$ & contain the zero contour \\
Grid-spacing ladder & nine values, 0.40--1.50\,mm & automatic resolution selection \\
Grid point budget & $6\times10^7$ & bound sampling cost \\
\midrule
\multicolumn{3}{@{}l}{\emph{Clean-up}} \\
Clip cap radius & $\max(1.8\,r,\ r+3\,\mathrm{tol}+1.0\,\mathrm{mm})$ & open ports by triangle removal \\
Taubin smoothing & 40 iterations, pass band 0.05 & boundary-preserving polish \\
Degenerate-triangle area & $\max(10^{-8},\,10^{-3}\mathrm{tol}^2)\,\mathrm{mm}^2$ & drop slivers \\
Hole-fill cap & $0.8\rho=0.56$\,mm & close slits below port scale \\
\midrule
\multicolumn{3}{@{}l}{\emph{Acceptance}} \\
Port tolerance $t_p$ & $\max(1.0\,r_p,\,1.8\,\mathrm{mm})$ & port--loop matching \\
Interface assertions & centre $<10^{-4}$\,mm; normals $<-0.95$ & shared-port consistency \\
\bottomrule
\end{tabular}
\end{center}

\subsection*{SM1. Datasets and case selection}
\phantomsection\label{sec:sm1}

ATM22~\cite{zhang2023multisite} provides non-contrast chest CT scans with expert airway labels from multiple sites,
ranging from healthy subjects to COVID-19 patients; 15 cases from its training set, spanning small and large airway
trees, were used.
AeroPath~\cite{stoverud2024aeropath} provides 27 contrast-enhanced CT angiography scans from patients investigated for
lung cancer, with pathologies including malignant tumours, sarcoidosis and emphysema, and anatomical consequences such
as an enlarged trachea, collapsed lung segments and tumour-occluded airways; all 27 were used. Its annotations contain
fewer branches than those of ATM22 ($128\pm56$ against $208\pm74$), reflecting airways obscured or occluded by disease.
In both datasets the pipeline input was the distributed reference annotation rather than an automatic segmentation, so
the results isolate the reconstruction stage from segmentation error.

Median in-plane voxel spacing was 0.781\,mm for ATM22 (range 0.734--0.904\,mm) and 0.703\,mm for AeroPath (range
0.563--0.824\,mm). Slice spacing was 0.500\,mm in every ATM22 case, finer than the in-plane spacing, and ranged from
0.499 to 1.250\,mm in AeroPath, where it exceeded the in-plane spacing in some scans (largest slice-to-in-plane ratio
1.778). All 42 scans had an identity direction matrix. The pipeline applies the full image affine, so anisotropic and
obliquely acquired data are handled by construction; the cohorts exercise moderate anisotropy but no oblique
acquisition and no slice thicker than 1.25\,mm.

Parameter values (Table~\ref{tab:params} and Table~\ref{tab:params-full}) were fixed before the evaluation run and are
recorded in the per-case pruning and validation records; the ablation of Section~\ref{sec:res-ablation} characterises
their effect rather than selecting them. Both datasets are publicly available and de-identified, and their use required
no further ethical approval.

\subsection*{SM2. Coordinates and graph construction}
\phantomsection\label{sec:sm2}

\paragraph{Coordinate conventions}
Two conventions are used and kept distinct. Voxel topology --- connectivity, skeletonisation and the radius
transform --- is computed in array index order $(z,y,x)$ with the array-order spacing $(s_z,s_y,s_x)$. Geometry is
expressed in physical coordinates in millimetres through the full affine of Eq.~\eqref{eq:affine}. Because origin,
spacing and direction matrix are all applied, anisotropic and oblique volumes are handled without resampling, and
voxel anisotropy is accounted for analytically wherever a physical length is computed. Coordinates follow the ITK LPS
convention throughout; exported surfaces are not converted to RAS.

\paragraph{Mask preprocessing}
The largest-component filter removes isolated false-positive fragments while preserving the tracheobronchial tree; in
13 of the 42 cases it removed at least one fragment. No morphological closing, hole filling or dilation is applied:
the wall is reconstructed from the centreline and radius field rather than from the mask boundary, so the mask serves
as anatomical evidence and is not itself smoothed. The 18-connectivity of the component filter and the 26-connectivity
of the skeleton analysis differ deliberately: the stricter rule avoids joining fragments that touch only at a corner,
while the looser rule follows one-voxel-wide chains that are diagonal on the lattice.

\paragraph{Radius}
The distance transform measures the distance from a foreground voxel centre to the nearest background voxel centre,
with no half-voxel correction. The radius is therefore the inscribed radius of the lumen, which the reconstructed wall
inherits; its consequence for lumen calibre is quantified in Section~\ref{sec:res-fidelity}.

\paragraph{Graph}
Skeleton voxels with a 26-neighbour degree other than two are critical: endpoints at degree one and junctions at
degree three or more. Spatially adjacent critical voxels are clustered into a single node, positioned at the mean of
its cluster in world coordinates.

\paragraph{Rooting and extent}
In the root key $(r_{\max},r_{\mathrm{mean}},z)$ the axial coordinate only breaks ties, so the root is the endpoint
with the largest local airway radius, the tracheal inlet. Pruning repeats until no terminal edge qualifies, after which
the tree contains only the root, branch nodes with two or more children and outlets. Depth is counted in branch nodes
on the collapsed tree rather than in graph hops: hop count depends on how finely each skeleton happens to be
discretised, and in our cohorts the maximum hop depth ranged from 14 to 23 across cases of comparable size, so a fixed
hop cut-off would truncate some trees far earlier than others. Thirty-four of the 42 cases reached the generation-8 cap;
the remainder terminated at generation 6 or 7, where the segmented tree ended before the cap would bind.

\subsection*{SM3. Shared ports and semantic blocks}
\phantomsection\label{sec:sm3}

\paragraph{Port orientation}
An inlet port uses the negated start tangent and an outlet port the end tangent, so that mating blocks oppose. A
junction-cut port uses the chord direction from the junction centre to the cut point rather than the local tangent,
which keeps the pipe cut and the junction cut coplanar. Because a block stores only a port identifier, the pipe on an
edge and the junction at its end read the same object; interface consistency is a property of the representation, not
a tolerance reached by fitting or repair, and is checked as a software assertion rather than reported as a result.

\paragraph{Block census}
Across the 42 cases the decomposition produced 6191 blocks: 4153 curved pipes, 1968 Y2 bifurcations, 60 Y3
trifurcations in 29 cases, 9 straight pipes and a single general junction.

\paragraph{Primitive construction}
The semantic assembly is a single PolyData carrying a block index and a block-type code per cell. Curved pipes are
swept tubes with 96 circumferential samples per ring on the smoothed profiles of the fused wall; the sweeping frame is
propagated by rotation-minimising parallel transport, and the two endpoint tangents are replaced by the mating port
normals so that consecutive blocks remain coplanar. Y2 and Y3 blocks are zero isosurfaces of the hard minimum of
constant-radius tube fields about sampled parent and daughter centrelines, contoured on a local Cartesian grid and
transformed into world coordinates. Daughter orientations are expressed as spherical angles in a frame whose local
$z$ axis points from the parent port to the junction centre, and daughter centrelines are straight. Port positions and
normals of these blocks are analytical metadata, not quantities recovered from the contoured surface. A fallback path
builds a trifurcation from two overlapping Y2 primitives when the native construction fails; it was not required for
any of the 60 Y3 blocks. The single general junction in the cohort is represented only in the fused field.

\subsection*{SM4. Fused-wall construction}
\phantomsection\label{sec:sm4}

\paragraph{Smoothing scale}
Because the Gaussian kernel widths are proportional to branch length, the radius kernel of a 60\,mm branch spans
about 11\,mm, and a narrowing shorter than that is attenuated. This bounds the shortest lesion the fused wall can
represent faithfully, and is why the pathology study scales its lesion width with branch length. The curved semantic
pipes use the same smoothed profiles.

\paragraph{Radius floor}
The semantic assembly retains the unfloored radii, so the effect of the floor is confined to the fused wall; it is
quantified in Sections~\ref{sec:res-fidelity} and~\ref{sec:res-ablation}.

\paragraph{Hard union and narrow band}
The hard minimum sets each branch's radius by its own primitive rather than inflating it at junctions, as a smooth
union would. Where two branches run close together the minimum can be attained by a neighbouring branch's primitive,
so the wall there follows the neighbour's envelope (Section~\ref{sec:res-fidelity}). The field is evaluated only
inside a narrow band around the centrelines, of half-width $r_{\max}+\tfrac12\ell_{\mathrm{seg}}^{\max}+3h$, found by
a nearest-midpoint query on a $k$-d tree; outside the band the field cannot reach the zero contour, so the restriction
does not alter the extracted surface.

\paragraph{Grid and extraction}
The grid spans the primitive bounding box plus a margin of $\max(2r_{\max}+3h,\,8\,\mathrm{mm})$. The ladder has nine
values from 0.40 to 1.50\,mm; if none satisfies the point budget the run is reported as failed rather than continuing
at a coarser spacing. Because the grid carries physical spacing the surface is already in millimetres, and because the
field is negative inside, triangle winding needs no separate reorientation.

\subsection*{SM5. Clipping, repair and acceptance}
\phantomsection\label{sec:sm5}

\paragraph{Clipping and component selection}
Ports are opened by triangle removal rather than a Boolean cut, within a cap of radius
$\max(1.8\,r,\ r+3\,\mathrm{tol}+1.0\,\mathrm{mm})$. After clipping every component containing an expected port is
retained, not only the largest. Keeping only the largest would risk removing correct anatomy: where a large-radius
port's cap reaches across to a neighbouring branch, the clip can sever the neck joining a sub-tree to the trunk.
Retaining every port-bearing component means such a case produces more than one component and fails acceptance rather
than silently losing anatomy. The selection is repeated after repair, because repair can detach small fragments.

\paragraph{Smoothing and repair}
Taubin smoothing is preferred to Laplacian smoothing because it avoids volumetric shrinkage, and boundary and feature
smoothing are disabled so that the open port rings, and hence the boundary loops, do not move. Degenerate triangles
below an area of $\max(10^{-8},\,10^{-3}\mathrm{tol}^2)\,\mathrm{mm}^2$ are dropped, and non-manifold edges are
repaired by iteratively removing the smallest-area offending triangle until every edge is incident to at most two
faces. The hole fill is capped at $0.8\rho=0.56$\,mm, below the smallest physical port loop, so it closes residual
slits but never a port.

\paragraph{Matching}
Optimal one-to-one assignment makes the result independent of the order in which ports are listed. Inadmissible
pairings are rejected rather than absorbed: without rejection, a global assignment facing too few loops reshuffles
otherwise correct matches to minimise total cost, reporting large errors on ports whose own geometry is sound. The
absolute term of Eq.~\eqref{eq:tolerance} reflects a displacement between the nominal port plane and the realised
surface that is approximately constant in millimetres rather than proportional to radius; re-evaluating the field at
one third of the production grid spacing does not reduce it, so it is a small geometric disagreement, not a sampling
limit. The term is fixed across all cases and cohorts. As a sensitivity figure, acceptance is also reported under a
stricter criterion that additionally requires $|\bar r_{\mathrm{loop}}-r_p|<0.5\,r_p$.

\subsection*{SM6. Controlled pathology editing}
\phantomsection\label{sec:sm6}

\paragraph{Cohort and sites}
Variants were generated from five reconstructions (ATM\_003, ATM\_045, AeroPath\_05, AeroPath\_20 and AeroPath\_21)
at two proximal sites each: the trachea, taken as the root's first child edge, and a main-bronchus branch selected as
the carina child most aligned with the tracheal chord. Because that rule is geometric, the branch is reported as the
algorithmically selected main-bronchus branch without a claim of verified laterality; one case carries an ambiguity
flag under the rule. All 40 variants and the five healthy baselines used the same frozen production settings, with no
case-specific tuning. Severity denotes fractional radius change rather than area reduction: for a circular section the
expected area ratios are 0.5625, 0.25, 0.0625 and 2.25. The first and last radius samples of the edited primitive are
restored to their healthy values, the radius floor is applied as usual and the centreline is unchanged, so the edit
alters calibre without displacing the branch.

\paragraph{Severity and locality}
Control is measured on the final fused surfaces rather than on the prescribed radius arrays. Healthy and edited meshes
are sampled at matched stations along the healthy branch centreline and cut by the plane orthogonal to the local
tangent. The primary measurement is the ratio of edited to healthy equivalent radius at $u=0.45$, using the same station
in both geometries; the area ratio is defined likewise. Locality is assessed outside the interval $0.20\le u\le0.70$,
fixed from the prescribed Gaussian parameters before any reconstructed surface was inspected. The primary analysis is
restricted to the branch shaft by a margin of $\max(2.5\,\mathrm{mm},\,1.2\,r_{\mathrm{healthy}})$ that excludes
endpoint and junction neighbourhoods, with an unrestricted analysis retained as a sensitivity check.

\paragraph{Boundary loops}
Canonical ports are reused unchanged by the editing workflow, so healthy and edited loops are paired by port identity
and compared by the shift of their realised centroids. This measures variation between independently fused surfaces,
not movement of the ports.

\subsection*{SM7. Evaluation protocol}
\phantomsection\label{sec:sm7}

\paragraph{Baselines}
B1 applies marching cubes at iso-level 0.5 to the cleaned largest component of the mask. B2 applies Taubin smoothing to
B1 with the settings used for the fused wall. B3 builds a swept tube with 24 circumferential samples per ring on each
pruned graph edge independently, using the unsmoothed graph profiles, and concatenates them without union or trimming,
so capped tube ends overlap at junctions by construction. Topology metrics are computed by the same functions for every
strategy. No composite readiness score is reported, because the capability differences are categorical.

\paragraph{Surface-to-mask agreement}
Vertices are subsampled to at most $10^5$ per surface under a fixed seed. The inside-mask fraction is obtained by
trilinear interpolation of the mask at the same vertices. Because the reconstruction is a regenerated parametric wall
rather than a voxel-faithful copy, these quantities measure anatomical agreement rather than surface replication, and a
mask-derived surface is nearer the mask by construction.

\paragraph{Lumen calibre}
Both mask and wall are measured with the same star-shaped estimator about the station centre (180 rays, first boundary
crossing) and area-equivalent radius (Eq.~\eqref{eq:s-req}), rather than by connected-component area, because in proximal generations the in-plane mask component can
merge with a touching neighbouring airway and return radii several times the true value. The plane normal is the first
principal direction of the raw centreline over a window of $\pm\max(2r,\,2\,\mathrm{mm})$: centrelines are
voxel-stepped at about 0.5\,mm, so a short-window tangent can differ from the longer-window one by tens of degrees and
inflate a cross-section by $1/\cos\theta$. Taking the window on the raw centreline, not the reconstruction's own axis,
keeps the measurement plane independent of the geometry under test. Stations are stratified, not excluded, by whether
another branch's floored wall comes within the station's own radius.

\paragraph{Branching statistics}
Branch angle is the angle between parent and child tangents, each measured over a fractional arc-length window near the
junction. Branch length is the chord between successive branch points, matching the comparison literature, and
branches separated by a bifurcation that lost a child to pruning are kept distinct. Strahler orders are computed with
branches collapsed by union--find, incrementing only when the two highest child orders tie; Horsfield orders increment
at every segment.

\paragraph{Ablation and cost}
A setting is reported as failing when the resulting wall does not meet the acceptance criteria. Runtime and peak
resident memory are the median of three repeats per case.

\paragraph{Statistics}
Each case contributes many stations, so pooled station statistics are not independent. Limits of agreement use the
Bland--Altman method for repeated measurements per subject~\cite{blandaltman2007}, with case-level medians and
interquartile ranges reported alongside; almost all the variance of the differences is within-case, so the corrected
and naive standard deviations agree to three decimal places. Continuous quantities are given as mean\,$\pm$\,SD across
cases, or median [IQR] where skewed. The 40 pathology variants are repeated interventions on five cases rather than 40
independent subjects and are summarised accordingly.

\subsection*{SM8. Downstream meshing and OpenFOAM simulation}
\phantomsection\label{sec:sm8}

\paragraph{Geometry transfer and meshing}
The fused wall was exported as STL in millimetres in the LPS frame. Its inlet loop and 65 outlet loops were closed with
cap surfaces, giving a watertight boundary of 243\,620 triangles. The ICEM CFD Robust (Octree) mesher used an element scale factor of 1.3, a maximum
element size of 2.0\,mm and curvature- and proximity-based refinement (minimum size 0.5\,mm, 30 elements per
$360^\circ$ of curvature, one element across gaps), followed by smoothing. The mesh was written in ANSYS mesh format and
converted to OpenFOAM by a script that splits the outlet caps into one patch per connected opening. The three openings
treated as wall had equivalent diameters of 0.37--0.68\,mm, below the in-plane voxel spacing. Transfer fidelity was
measured by the distance from every boundary node of the volume mesh to the STL surface and by comparing boundary area
and enclosed volume.

\paragraph{Inlet condition}
The inflow was imposed on the inlet cap ($A=59.7\,\mathrm{mm}^2$) with the profile $u\propto(d/d_{\max})(2-d/d_{\max})$,
where $d$ is the distance from the rim; its peak-to-mean ratio is 1.75. With the mean velocity $U=1.67$\,m\,s$^{-1}$ and
the inlet hydraulic diameter $D_h=8.71$\,mm, the inlet Reynolds number $Re=\rho_f U D_h/\mu$ is about 960, for which
steady laminar flow is a reasonable assumption.

\paragraph{Outlet resistances}
For an outlet of equivalent diameter $D_k$, the generation $g$ is the Weibel generation whose diameter $D_g$ is closest
on a logarithmic scale, and the missing subtree runs from generation $g+1$ to 16 with every tube scaled by $s=D_k/D_g$
and dimensions from Weibel's model~A. Resistances are combined from generation 16 upwards, two daughters in parallel
feeding each parent in series, and the outlet feeds two generation-$(g+1)$ branches in parallel. The 62 outlets fell in
generations 5--12, with $R_k$ from $2.3\times10^5$ to $4.8\times10^6$\,Pa\,s\,m$^{-3}$; one generation more or less
changes a resistance by about $-15$\,\% or $+17$\,\%. Pressures are gauge values relative to the distal end of the
modelled subtrees.

\paragraph{Coupling and numerics}
Because $Q_k$ is not known in advance, Eq.~\eqref{eq:outlet-resistance} was enforced iteratively: outlet pressures were
held fixed for 20 solver iterations, the outlet flows were read, and each $p_k$ was updated by a chord-Newton step with
a diagonal estimate of the response of $Q_k$ to $p_k$. Convection used bounded \texttt{linearUpwind}, gradients were
linear (cell-limited for velocity) and non-orthogonal corrections were limited. Pressure was solved with GAMG and
velocity with a symmetric Gauss--Seidel smoother, with a velocity relaxation factor of 0.8. The field was initialised
from a solution with zero outlet pressure. The boundary-condition error is $\max_k|p_k-R_kQ_k|/\max_k R_kQ_k$.

\paragraph{Post-processing}
All quantities were computed directly from the OpenFOAM mesh and field files. Outlet flows are sums of the stored face
fluxes. To split the flow between the lungs, the cells within 12\,mm of the carina were removed and the remaining cell
graph was divided into connected components; the result did not change for cut radii of 12--25\,mm. The flow share
predicted from the outlet resistances alone, $(1/R_k)/\sum_j(1/R_j)$, is the reference for the computed shares. In the
whole-airway speed view each pixel shows the largest cell speed along the viewing direction.

Sections S1--S3 are planes normal to the local airway axis through the point 10\,mm from the carina along that axis, in
the trachea (S1) and in the $-x$ (S2) and $+x$ (S3) main bronchi; $\pm x$ are CT image axes. The carina is the cell of
highest pressure, on the carinal ridge. Each axis was first estimated from the branch centroids 16 and 25\,mm from the
carina, where the three branches separate, and then refined from the centroids of sections 2.5\,mm on either side of the
plane. On each section the velocity is split into $u_{\mathrm{ax}}=\mathbf{u}\cdot\mathbf{n}$ and
$\mathbf{u}_{\mathrm{tr}}=\mathbf{u}-u_{\mathrm{ax}}\mathbf{n}$. Quoted section quantities are area-weighted cell values,
which reproduce the solver's branch flows to within 0.5\,\%; contour plots interpolate node values (the mean of the
adjacent cells, zero on the wall) linearly within each cut cell, without smoothing. Streamlines are integrated with a
second-order Runge--Kutta scheme through the same interpolation, from inlet seeds placed in proportion to the local
inflow.

\subsection*{SM9. Outputs and reproducibility}
\phantomsection\label{sec:sm9}

Each run writes the semantic assembly and the fused wall as VTK PolyData and STL, the semantic PolyData carrying a
per-cell block index and block-type code that the STL format cannot preserve. The external port manifest, written as
CSV and JSON, lists for each port its identifier, kind, centre, normal, radius and a boundary label of the form
\texttt{inlet}, \texttt{outlet\_01} and so on; these labels are structural rather than anatomical, as no lobar or
segmental names are assigned. Further records cover the shared-port graph including internal interfaces, the block
decomposition and the full validation report with per-port errors and the acceptance verdict. All geometry is in
millimetres in the ITK LPS convention. The reconstruction uses no random state; evaluation scripts that sample randomly
use recorded seeds. Graph construction is custom and uses no external graph library, and exact package versions are
pinned in the released environment specification. Reading ATM22 through the supplied loaders requires a small
compatibility layer that maps ATM22 identifiers and filenames onto the AeroPath directory convention; the script that
creates it is included.

\subsection*{SM10. Quantities shown in the figures}
\phantomsection\label{sec:sm10}
\setcounter{equation}{0}
\makeatletter
\g@addto@macro\normalsize{\abovedisplayskip=4pt plus 1pt\belowdisplayskip=4pt plus 1pt\abovedisplayshortskip=1pt\belowdisplayshortskip=2pt}
\makeatother\normalsize
\renewcommand{\theequation}{S\arabic{equation}}

This section defines every derived quantity plotted in the figures, grouped by figure. Figure~\ref{fig:method_panels}
needs no further definitions: its panels illustrate Eqs.~\eqref{eq:affine}, \eqref{eq:frame},
\eqref{eq:capsule-t}--\eqref{eq:union} and~\eqref{eq:tolerance} of the main text. The equivalent radius of any planar cross-section of area $A$, used for the lumen-calibre, pathology and CFD-section
measurements, is
\begin{equation}
  r_{\mathrm{eq}}=\sqrt{A/\pi}.
  \label{eq:s-req}
\end{equation}

\paragraph{Pathology control (Fig.~\ref{fig:pathology-quant}, Fig.~\ref{fig:pathology-errors-supp})}
For each variant $v$ the achieved radius ratio, plotted against the commanded
ratio $q_r^{\mathrm{cmd}}$ in Fig.~\ref{fig:pathology-quant}, and its error are
\begin{equation}
  q_{r,v}^{\mathrm{ach}}=\frac{r^{\mathrm{edit}}_{\mathrm{eq}}(c)}{r^{\mathrm{healthy}}_{\mathrm{eq}}(c)},\qquad
  E_{r,v}=\bigl|q_{r,v}^{\mathrm{ach}}-q_{r,v}^{\mathrm{cmd}}\bigr|,\qquad
  \mathrm{MAE}=\frac{1}{40}\sum_{v=1}^{40}E_{r,v},
  \label{eq:s-qr}
\end{equation}
with $c=0.45$ and the same station in both geometries; $E_r$ is plotted in Fig.~\ref{fig:pathology-errors-supp}a.
Locality (Fig.~\ref{fig:pathology-errors-supp}b) is the 95th percentile of the radius deviation over the set $U_v$ of
shaft stations outside the lesion interval $0.20\le u\le0.70$ and outside the endpoint and junction margins,
\begin{equation}
  \Delta_{\mathrm{shaft},v}=\operatorname*{P95}_{u\in U_v}\bigl|r^{\mathrm{edit}}_{\mathrm{eq}}(u)-r^{\mathrm{healthy}}_{\mathrm{eq}}(u)\bigr|.
  \label{eq:s-shaft}
\end{equation}
Boundary-loop stability (Fig.~\ref{fig:pathology-errors-supp}c) is the 95th percentile, over the external ports $p$ of a
variant, of the shift of the realised loop centroid $\bar{\mathbf c}_p$ between the healthy and edited surfaces,
\begin{equation}
  \Delta_{\mathrm{loop},v}=\operatorname*{P95}_{p}\bigl\|\bar{\mathbf c}^{\,\mathrm{edit}}_p-\bar{\mathbf c}^{\,\mathrm{healthy}}_p\bigr\|_2 .
  \label{eq:s-loop}
\end{equation}

\paragraph{Volume mesh (Fig.~\ref{fig:mesh}, Table~\ref{tab:cfd-setup})}
For an internal face $f$ with area vector $\mathbf S_f$, joining cells whose centres are separated by $\mathbf d_f$, the
non-orthogonality and skewness reported by \texttt{checkMesh} are
\begin{equation}
  \theta_f=\arccos\frac{\mathbf d_f\cdot\mathbf S_f}{|\mathbf d_f|\,|\mathbf S_f|},\qquad
  \mathrm{sk}_f=\frac{|\mathbf x_f-\mathbf x_f^{\ast}|}{|\mathbf d_f|},
  \label{eq:s-mesh}
\end{equation}
where $\mathbf x_f$ is the face centre and $\mathbf x_f^{\ast}$ the point where the line between the two cell centres
crosses the face plane.

\paragraph{Flow field (Fig.~\ref{fig:cfd})}
Panel~a shows, for each pixel $(x,z)$ of a 0.3\,mm grid in the viewing plane, the largest cell speed along the viewing
direction $y$,
\begin{equation}
  I(x,z)=\max_{i\,:\,\mathbf x_i\in\Pi(x,z)}|\mathbf u_i|,
  \label{eq:s-mip}
\end{equation}
where $\Pi(x,z)$ is the column of space projecting onto the pixel and $\mathbf x_i$ the cell centroid; pixels inside the
airway silhouette that contain no centroid take the largest value of their eight neighbours. Panel~b shows the wall
static pressure $p$, whose zero is the distal end of the modelled subtrees through Eq.~\eqref{eq:outlet-resistance}.
Panel~c compares the computed outlet flow share with the share predicted from the outlet resistances alone,
\begin{equation}
  s_k=100\,\frac{Q_k}{Q_{\mathrm{in}}},\qquad
  \hat s_k=100\,\frac{R_k^{-1}}{\sum_{j}R_j^{-1}},
  \label{eq:s-share}
\end{equation}
and reports their Pearson correlation; $\hat s_k$ is the split that would arise if the reconstructed airways had no
resistance. Panel~d shows sections $\{\mathbf x:(\mathbf x-\mathbf c)\cdot\mathbf n=0\}$ normal to the local airway axis
$\mathbf n$, on which the velocity is decomposed as
\begin{equation}
  u_{\mathrm{ax}}=\mathbf u\cdot\mathbf n,\qquad
  \mathbf u_{\mathrm{tr}}=\mathbf u-u_{\mathrm{ax}}\mathbf n ,
  \label{eq:s-decomp}
\end{equation}
the colour showing $u_{\mathrm{ax}}$ (upper row) and $|\mathbf u_{\mathrm{tr}}|$ (lower row), and the arrows the
in-plane components of $\mathbf u_{\mathrm{tr}}$. The displacement of the axial maximum quoted in
Section~\ref{sec:res-cfd} is the distance between the area-weighted centroid of the fastest 10\,\% of the section and the
section centroid.

\paragraph{Model and convergence (Figs.~\ref{fig:cfd-model} and~\ref{fig:cfd-conv})}
Each outlet is assigned the Weibel generation whose diameter is closest on a logarithmic scale, and its resistance is
accumulated from generation 16 upwards, with $R_j$ from Eq.~\eqref{eq:weibel-resistance}:
\begin{equation}
  g_k=\arg\min_{g}\bigl|\ln D_k-\ln D_g\bigr|,\qquad
  R^{(16)}=R_{16},\quad R^{(j)}=R_j+\tfrac12R^{(j+1)},\quad R_k=\tfrac12R^{(g_k+1)} .
  \label{eq:s-resistance}
\end{equation}
The grey bars of Fig.~\ref{fig:cfd-model}b repeat Eq.~\eqref{eq:s-resistance} with $g_k\pm1$. The inlet profile of
Fig.~\ref{fig:cfd-model}c is
\begin{equation}
  u_n(d)=U_0\,\frac{d}{d_{\max}}\Bigl(2-\frac{d}{d_{\max}}\Bigr),\qquad
  U_0:\ \sum_f u_n(d_f)\,A_f=Q,
  \label{eq:s-inlet}
\end{equation}
where $d$ is the distance from the rim of the inlet cap and $A_f$ the area of inlet face $f$. Between outlet updates
$n-1$ and $n$, Fig.~\ref{fig:cfd-conv}a plots
\begin{equation}
  \varepsilon^{(n)}=\frac{\max_k\bigl|p_k-R_kQ_k\bigr|}{\max_kR_kQ_k},\qquad
  \Delta s^{(n)}=\max_k\bigl|s_k^{(n)}-s_k^{(n-1)}\bigr|,
  \label{eq:s-conv}
\end{equation}
and the outlet pressures are then updated by the chord-Newton step
$p_k\leftarrow p_k-(p_k-R_kQ_k)/(1+R_k/R_{\mathrm{loc},k})$, where $R_{\mathrm{loc},k}$ is the pressure-to-flow ratio of
outlet $k$ in the initial solution. Fig.~\ref{fig:cfd-conv}b plots the OpenFOAM scaled residual of each equation
$\mathbf A\boldsymbol\phi=\mathbf b$,
\begin{equation}
  \mathcal R=\frac{\sum\bigl|\mathbf b-\mathbf A\boldsymbol\phi\bigr|}
  {\sum\bigl(\bigl|\mathbf A\boldsymbol\phi-\mathbf A\bar{\boldsymbol\phi}\bigr|+\bigl|\mathbf b-\mathbf A\bar{\boldsymbol\phi}\bigr|\bigr)},
  \label{eq:s-residual}
\end{equation}
with $\bar{\boldsymbol\phi}$ the field average, evaluated at the start of each iteration.

\paragraph{Flow development and streamlines (Figs.~\ref{fig:cfd-streamlines} and~\ref{fig:cfd-profiles})}
On a section at distance $s$ from the carina, built from cut cells of area $A_i$, the plotted quantities are
\begin{equation}
  A(s)=\sum_iA_i,\qquad
  \bar u_{\mathrm{ax}}(s)=\frac{\sum_iA_i\,u_{\mathrm{ax},i}}{A(s)},\qquad
  \bar p(s)=\frac{\sum_iA_i\,p_i}{A(s)},\qquad
  \lambda(s)=\frac{\sum_iA_i\,|\mathbf u_{\mathrm{tr},i}|}{\sum_iA_i\,u_{\mathrm{ax},i}},
  \label{eq:s-profiles}
\end{equation}
where $\lambda$ measures the strength of the secondary flow. Streamlines solve
\begin{equation}
  d\mathbf x/d\sigma=\mathbf u(\mathbf x)/|\mathbf u(\mathbf x)|,\qquad \mathbf x(0)=\mathbf x_0,
  \label{eq:s-streamline}
\end{equation}
with arc length $\sigma$, integrated by a second-order Runge--Kutta scheme with $\Delta\sigma=0.25$\,mm (reduced near
the wall) through linear interpolation of node velocities within each tetrahedron. Seeds $\mathbf x_0$ are drawn on the
inlet with probability proportional to the face inflow $u_{n,f}A_f$.

\paragraph{Outlets (Fig.~\ref{fig:cfd-outlets})}
Panel~b fits $\log s_k=\log a+n\log D_k$ by least squares. Panel~c splits the pressure drop from the mean inlet pressure
$\bar p_{\mathrm{in}}$ to the distal end of each subtree into the parts within the reconstructed airways and within the
modelled subtree:
\begin{equation}
  \bar p_{\mathrm{in}}=(\bar p_{\mathrm{in}}-p_k)+p_k,\qquad f_k=(\bar p_{\mathrm{in}}-p_k)/\bar p_{\mathrm{in}},
  \label{eq:s-budget}
\end{equation}
the first term being the drop within the reconstructed airways and the second within the distal subtree.
Panel~d plots the cumulative share $C(m)=\sum_{k\le m}s_{(k)}$ of the outlets ordered by decreasing flow.

\clearpage

\section*{Supplementary Results 1: Reconstruction galleries}
\label{sec:supplementary-results-1}

The first two galleries show all 37 accepted reconstructions (14 ATM22 and
23 AeroPath cases). The final gallery shows the five cases that did not meet
the acceptance criteria, with annotated defects and close-up views.

\begin{center}
    \includegraphics[width=\textwidth,height=0.82\textheight,keepaspectratio]
        {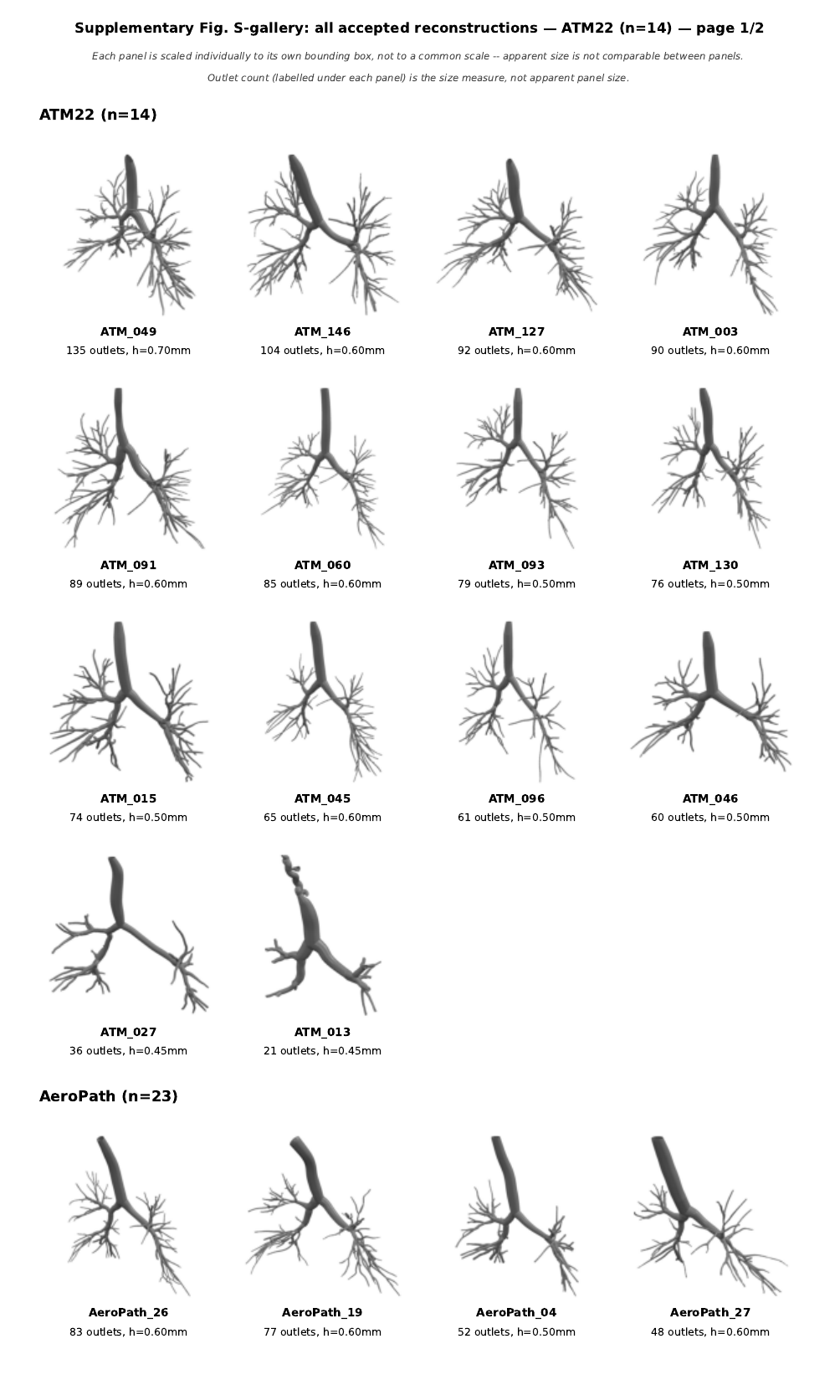}
\end{center}

\clearpage
\section*{Supplementary Results 1 (continued)}

\begin{center}
    \includegraphics[width=\textwidth,height=0.90\textheight,keepaspectratio]
        {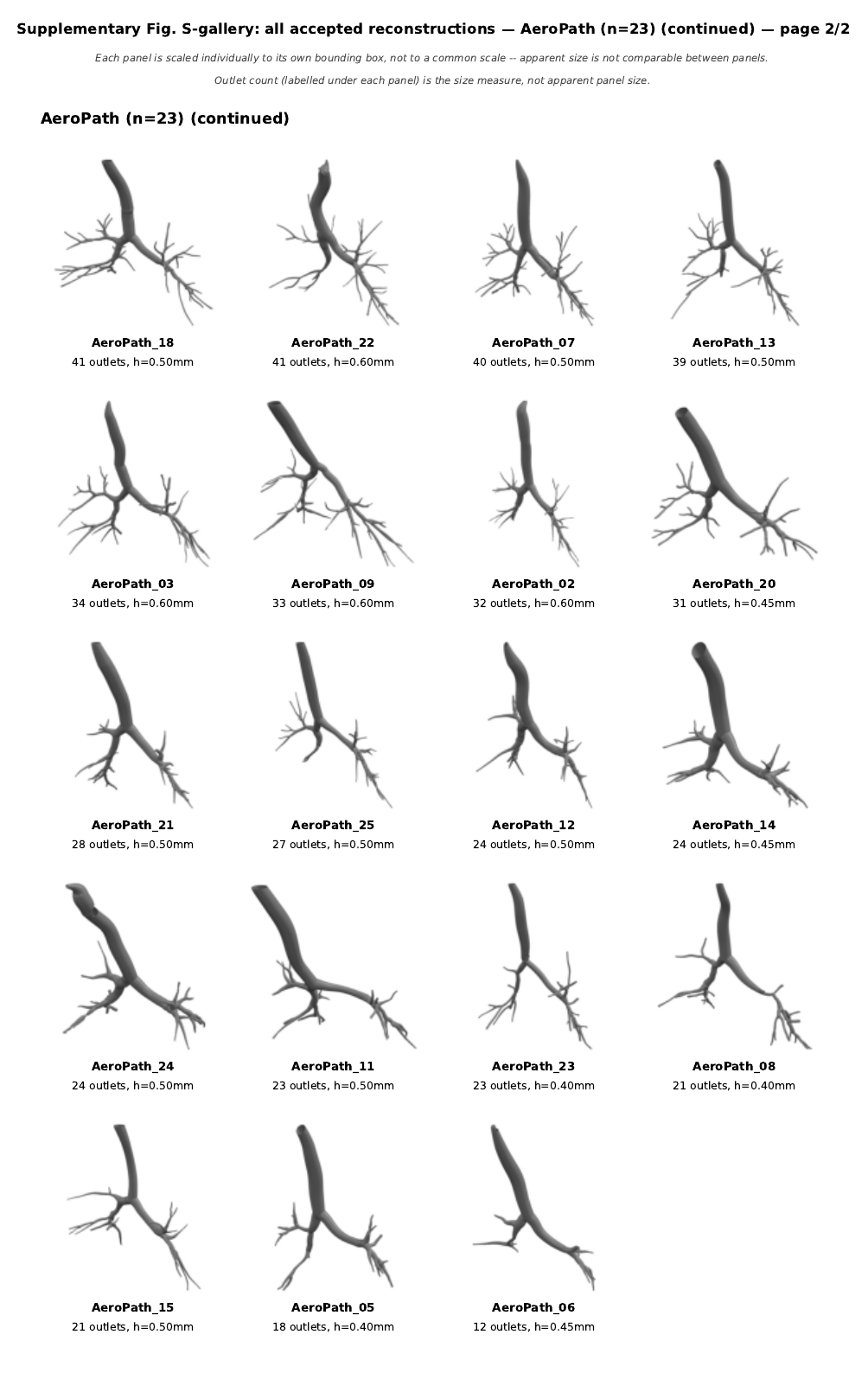}
\end{center}

\clearpage
\section*{Supplementary Results 1 (continued)}

\begin{center}
    \includegraphics[width=\textwidth,height=0.90\textheight,keepaspectratio]
        {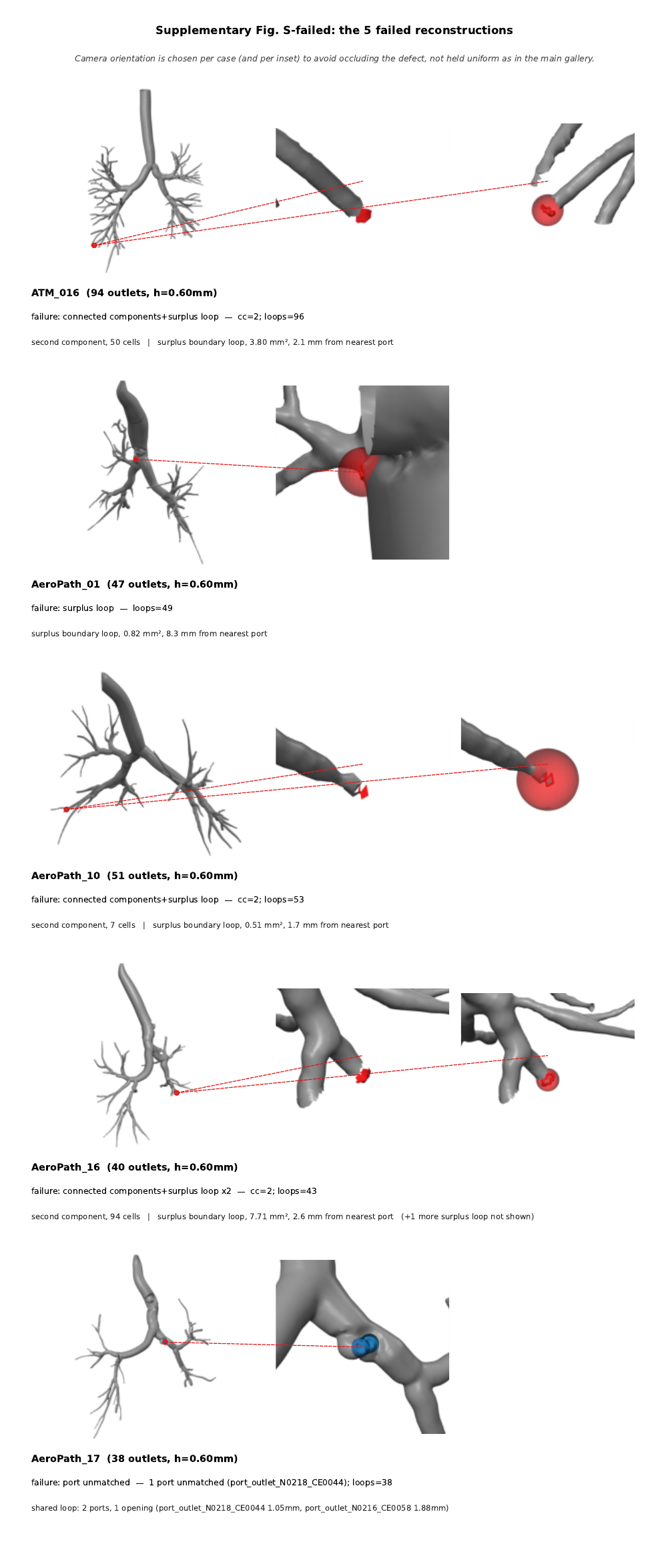}
\end{center}

\clearpage
\section*{Supplementary Results 2: Reconstruction and evaluation detail}
\label{sec:supplementary-results-2}

This section gives the case-level detail behind Sections~\ref{sec:res-acceptance}--\ref{sec:res-ablation}: every case
on common axes (Fig.~\ref{fig:s-percase}), the effect of the radius floor on port radii (Fig.~\ref{fig:s-port-radius}),
the branching statistics (Fig.~\ref{fig:s-branching}) and the computational cost and
self-intersection counts (Fig.~\ref{fig:s-cost}).

\begin{center}
    \includegraphics[width=\textwidth]{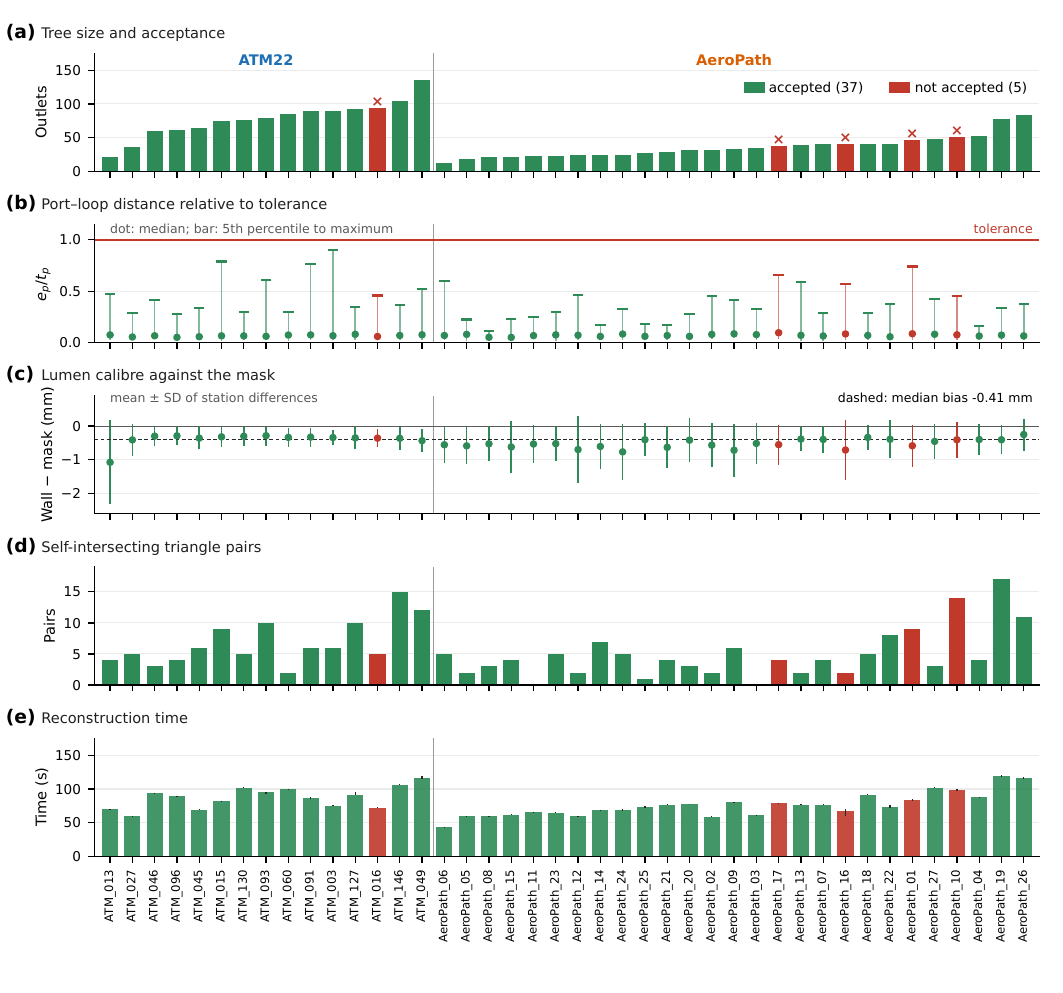}
    \captionof{figure}{Overview of all 42 reconstructions, ATM22 then AeroPath, each ordered by outlet count; green,
    accepted; red, not accepted. (a)~Number of outlets. (b)~Distance from each port to its matched loop relative to the
    port's tolerance: median (dot) and range from the 5th percentile to the maximum. (c)~Mean $\pm$ SD of the
    station-wise difference between wall and mask equivalent radius. (d)~Self-intersecting triangle pairs on the
    fused wall. (e)~Median reconstruction time over three repeats, with the range.}
    \label{fig:s-percase}
\end{center}

\clearpage
\begin{center}
    \includegraphics[width=\textwidth]{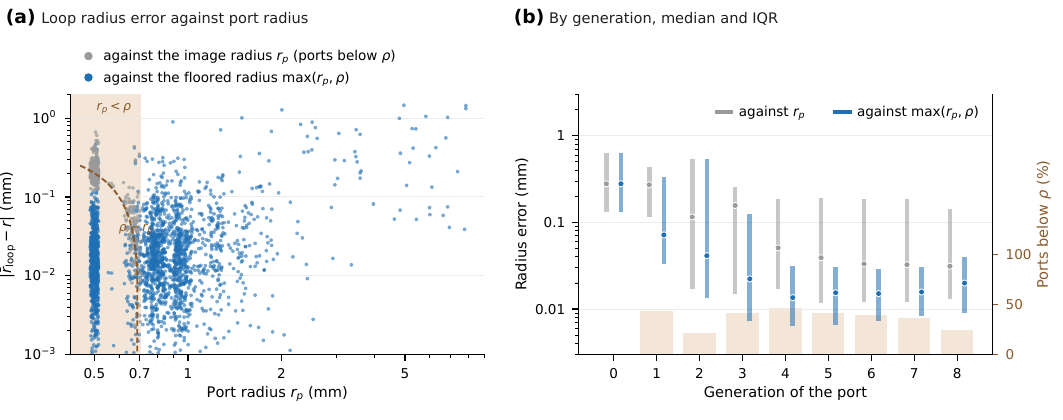}
    \captionof{figure}{Port radius error and the radius floor. (a)~Difference between the mean radius of each matched
    loop and the port radius (grey, shown for ports below the floor) or the floored radius $\max(r_p,\rho)$ (blue),
    against port radius; the shaded band marks ports below $\rho=0.7$\,mm, and the dashed curve the enlargement
    $\rho-r_p$. Measured against the floored radius, the median error of sub-floor ports falls from 0.195 to 0.015\,mm.
    (b)~The same two errors by port generation (median and interquartile range), with the share of ports below the floor
    in each generation.}
    \label{fig:s-port-radius}
\end{center}

\vspace{4mm}
\begin{center}
    \includegraphics[width=\textwidth]{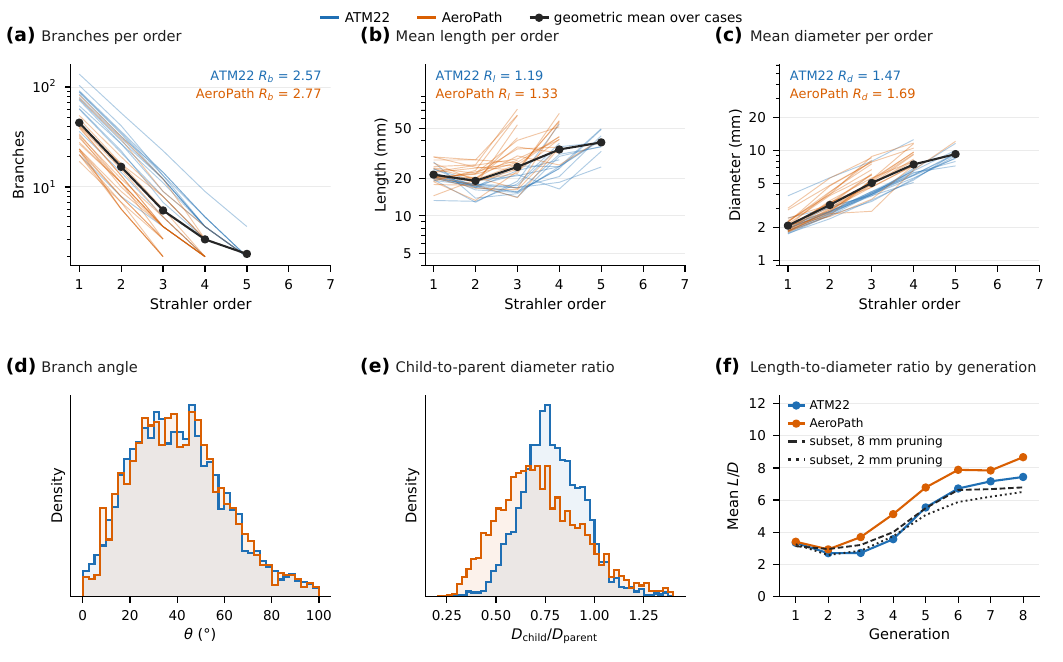}
    \captionof{figure}{Branching statistics of the 36 accepted cases with at least four Strahler orders. (a--c)~Number
    of branches, mean length and mean diameter against Strahler order for each case (thin lines) and their geometric
    mean (black), excluding the highest order as in the ratio fits; the mean ordering ratio of each cohort is given in
    the panel. (d)~Branch angle and (e)~child-to-parent diameter ratio, per branch. (f)~Mean length-to-diameter ratio by generation for each
    cohort, and for the twelve-case subset reconstructed with the terminal-pruning length threshold at 8\,mm and
    relaxed fourfold to 2\,mm.}
    \label{fig:s-branching}
\end{center}

\clearpage
\begin{center}
    \includegraphics[width=0.86\textwidth]{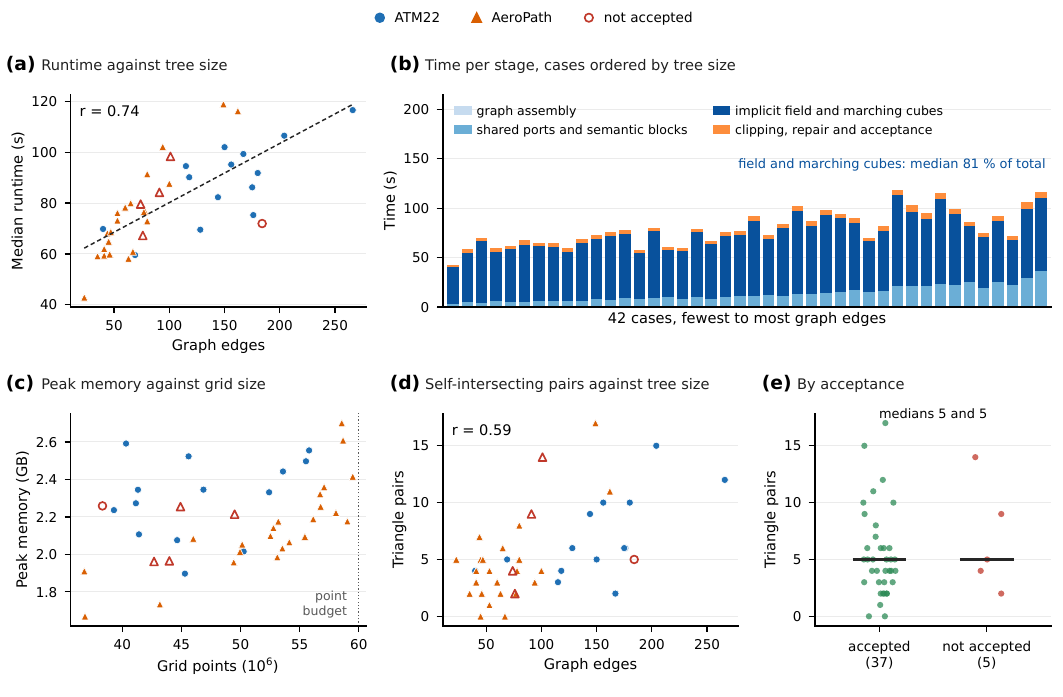}
    \captionof{figure}{Computational cost and self-intersections. (a)~Median reconstruction time against the number of
    graph edges, with a least-squares line; open red markers are cases that were not accepted. (b)~Time per pipeline
    stage for every case, ordered by tree size: graph assembly; shared ports and semantic blocks; implicit field and
    marching cubes; clipping, repair and acceptance. (c)~Peak resident memory against the number of grid points; the
    point budget is $6\times10^7$. (d)~Self-intersecting triangle pairs against the number of graph edges.
    (e)~Self-intersecting pairs in accepted and non-accepted cases. Eight threads on an Intel Xeon Gold 5218R CPU.}
    \label{fig:s-cost}
\end{center}

\vspace{3mm}
\section*{Supplementary Results 3: Pathology error distributions}
\label{sec:supplementary-results-pathology}

\begin{center}
    \includegraphics[width=0.9\textwidth]{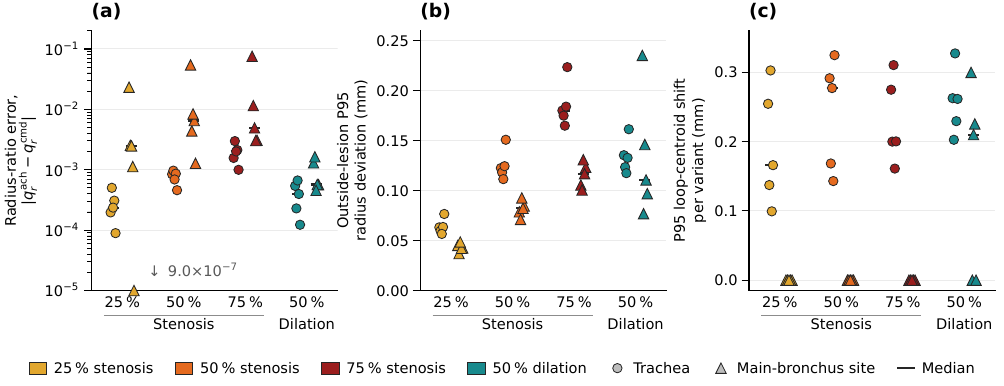}
    \captionof{figure}{Error distributions for the 40 pathology variants (5 cases $\times$ 2 sites $\times$ 4 edits);
    one point per variant, bars are group medians; circles, trachea; triangles, main-bronchus site; colour, prescribed
    edit. (a)~Absolute radius-ratio error at the lesion centre (Eq.~\eqref{eq:s-qr}), log axis; one value below the axis
    floor ($9.0\times10^{-7}$, AeroPath\_21) is drawn on the floor. (b)~P95 of the absolute equivalent-radius deviation
    along the branch shaft outside the lesion interval (Eq.~\eqref{eq:s-shaft}). (c)~P95 of the centroid shift of the
    realised boundary loops relative to the healthy surface (Eq.~\eqref{eq:s-loop}); the largest single-loop shift,
    0.411\,mm, exceeds the plotted maximum. Loops move only when an edit changes the largest primitive radius, which
    sets the extraction grid; the 17 main-bronchus variants at 0\,mm are measured zeros.}
    \label{fig:pathology-errors-supp}
\end{center}

\clearpage
\section*{Supplementary Results 4: Downstream CFD demonstration}
\label{sec:supplementary-results-cfd}
\renewcommand{\thetable}{S\arabic{table}}

This section documents the CFD demonstration of Section~\ref{sec:res-cfd} in full: the model and its boundary
conditions (Fig.~\ref{fig:cfd-model}), convergence (Fig.~\ref{fig:cfd-conv}), streamlines
(Fig.~\ref{fig:cfd-streamlines}), the development of the flow along the trachea and main bronchi
(Fig.~\ref{fig:cfd-profiles}) and the distribution of flow and pressure over the outlets (Fig.~\ref{fig:cfd-outlets},
Table~\ref{tab:cfd-outlets}). Every quantity is computed from the OpenFOAM mesh and field files of the converged
solution, steady laminar flow at 0.1\,L\,s$^{-1}$, by the scripts released with this material.

\paragraph{Model and boundary conditions}
The 62 outlets span equivalent diameters of 1.0--3.9\,mm and were assigned Weibel generations 5--12. Their resistances
fall from $4.8\times10^{6}$ to $2.3\times10^{5}$\,Pa\,s\,m$^{-3}$ as the diameter grows, close to a power law because
every subtree is a scaled copy of the same Weibel model (Fig.~\ref{fig:cfd-model}b). Assigning one generation more or
less would change a resistance by about $-15$\,\% or $+17$\,\%. The inlet profile has a mean normal velocity of
1.67\,m\,s$^{-1}$ and a peak-to-mean ratio of 1.75 over 176 faces (Fig.~\ref{fig:cfd-model}c).

\begin{center}
    \includegraphics[width=0.74\textwidth]{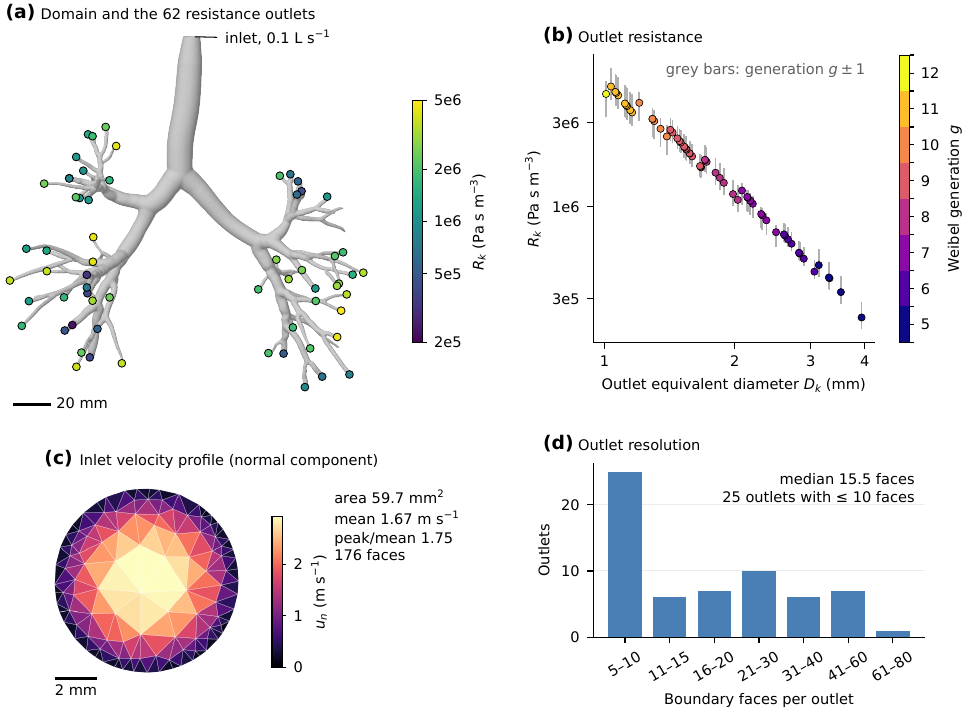}
    \captionof{figure}{CFD model and boundary conditions.
    (a)~Computational domain with the 62 outlets coloured by their resistance $R_k$; the three openings smaller than
    0.7\,mm are closed as wall and not shown.
    (b)~Outlet resistance against outlet equivalent diameter, coloured by the assigned Weibel generation $g$; grey bars
    span the resistances obtained with generation $g-1$ and $g+1$ (Eq.~\eqref{eq:s-resistance}).
    (c)~Normal component of the prescribed inlet velocity on the 176 inlet faces (Eq.~\eqref{eq:s-inlet}).
    (d)~Number of boundary faces on each outlet.}
    \label{fig:cfd-model}
\end{center}

\vspace{2mm}
\begin{center}
    \includegraphics[width=0.8\textwidth]{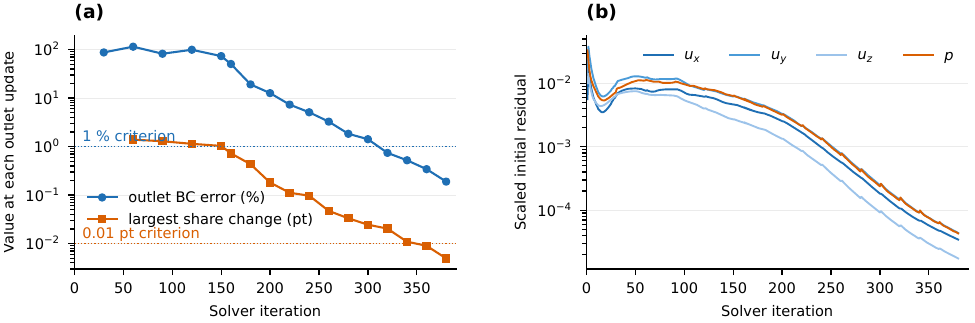}
    \captionof{figure}{Convergence of the solution.
    (a)~At each outlet update (every 20 solver iterations): the largest outlet boundary-condition error,
    $\max_k|p_k-R_kQ_k|/\max_k R_kQ_k$, and the largest change of any outlet flow share since the previous update (Eq.~\eqref{eq:s-conv}).
    Dotted lines mark the convergence criteria, which had to hold over two consecutive updates.
    (b)~Scaled initial residuals of the velocity components and pressure (Eq.~\eqref{eq:s-residual}).}
    \label{fig:cfd-conv}
\end{center}

\clearpage
\begin{center}
    \includegraphics[width=0.9\textwidth]{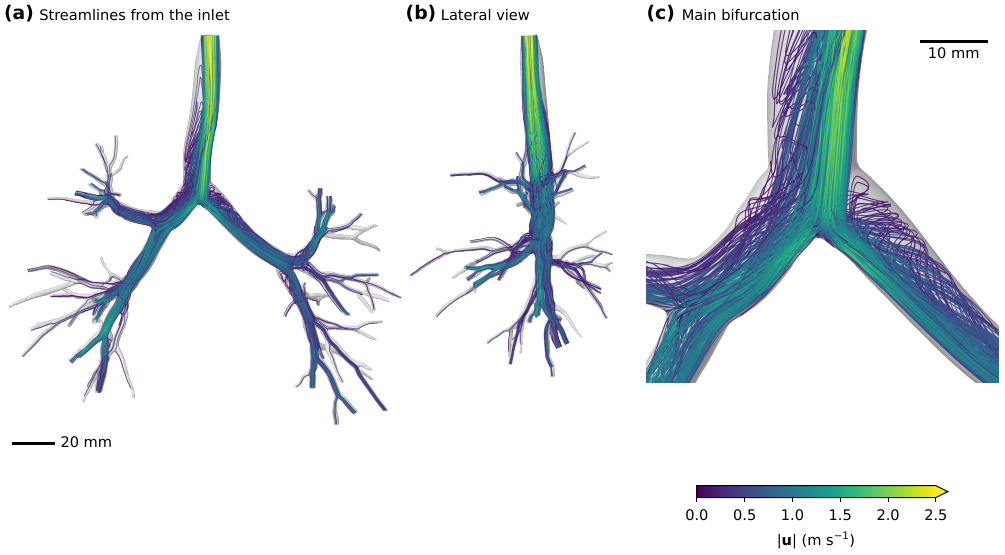}
    \captionof{figure}{Streamlines, coloured by speed. 600 streamlines start 0.3\,mm inside the inlet, seeded in
    proportion to the local inflow, and are integrated through the linearly interpolated velocity field
    (Eq.~\eqref{eq:s-streamline}).
    (a)~Front view. (b)~Lateral view. (c)~Close-up of the main bifurcation.}
    \label{fig:cfd-streamlines}
\end{center}

\vspace{4mm}
\begin{center}
\captionof{table}{Per-outlet results of the CFD demonstration at 0.1\,L\,s$^{-1}$, ordered by equivalent diameter
$D$. $g$, Weibel generation assigned to the outlet; $R_k$, outlet resistance ($10^{6}$\,Pa\,s\,m$^{-3}$); $p_k$,
outlet pressure relative to the distal end of the modelled subtree; $Q_k$, outlet flow; share, percentage of the
inflow.}
\label{tab:cfd-outlets}
\footnotesize
\begin{tabular}{@{}rrrrrrr@{\hspace{14pt}}rrrrrrr@{}}
\toprule
Outlet & $D$ (mm) & $g$ & $R_k$ & $p_k$ (Pa) & $Q_k$ (mL\,s$^{-1}$) & Share (\%) & Outlet & $D$ (mm) & $g$ & $R_k$ & $p_k$ (Pa) & $Q_k$ (mL\,s$^{-1}$) & Share (\%) \\
\midrule
01 & 3.93 & 5 & 0.23 & 1.78 & 7.607 & 7.61 & 32 & 1.67 & 9 & 1.67 & 1.61 & 0.966 & 0.97 \\
02 & 3.52 & 5 & 0.33 & 1.64 & 5.007 & 5.01 & 33 & 1.67 & 9 & 1.68 & 1.74 & 1.034 & 1.03 \\
03 & 3.32 & 5 & 0.39 & 1.86 & 4.757 & 4.76 & 34 & 1.66 & 9 & 1.70 & 1.12 & 0.657 & 0.66 \\
04 & 3.31 & 5 & 0.40 & 2.11 & 5.325 & 5.33 & 35 & 1.59 & 9 & 1.93 & 2.01 & 1.040 & 1.04 \\
05 & 3.13 & 5 & 0.46 & 1.52 & 3.269 & 3.27 & 36 & 1.57 & 9 & 2.00 & 1.30 & 0.649 & 0.65 \\
06 & 3.06 & 6 & 0.43 & 1.70 & 3.993 & 3.99 & 37 & 1.57 & 9 & 2.01 & 1.45 & 0.720 & 0.72 \\
07 & 2.89 & 6 & 0.51 & 2.30 & 4.552 & 4.55 & 38 & 1.55 & 9 & 2.11 & 1.47 & 0.696 & 0.70 \\
08 & 2.83 & 6 & 0.54 & 2.02 & 3.745 & 3.75 & 39 & 1.54 & 9 & 2.13 & 1.62 & 0.759 & 0.76 \\
09 & 2.82 & 6 & 0.55 & 1.92 & 3.504 & 3.50 & 40 & 1.53 & 9 & 2.18 & 1.45 & 0.668 & 0.67 \\
10 & 2.71 & 6 & 0.62 & 1.72 & 2.789 & 2.79 & 41 & 1.53 & 9 & 2.19 & 2.00 & 0.912 & 0.91 \\
11 & 2.66 & 6 & 0.65 & 1.69 & 2.591 & 2.59 & 42 & 1.50 & 9 & 2.32 & 1.54 & 0.663 & 0.66 \\
12 & 2.61 & 6 & 0.68 & 1.86 & 2.717 & 2.72 & 43 & 1.49 & 9 & 2.36 & 2.02 & 0.856 & 0.86 \\
13 & 2.60 & 6 & 0.70 & 1.70 & 2.439 & 2.44 & 44 & 1.47 & 9 & 2.44 & 1.41 & 0.576 & 0.58 \\
14 & 2.49 & 7 & 0.72 & 2.14 & 2.988 & 2.99 & 45 & 1.44 & 9 & 2.64 & 1.26 & 0.478 & 0.48 \\
15 & 2.37 & 7 & 0.84 & 1.39 & 1.664 & 1.66 & 46 & 1.42 & 9 & 2.73 & 2.13 & 0.778 & 0.78 \\
16 & 2.32 & 7 & 0.89 & 2.16 & 2.418 & 2.42 & 47 & 1.39 & 10 & 2.51 & 1.51 & 0.604 & 0.60 \\
17 & 2.30 & 7 & 0.91 & 1.45 & 1.598 & 1.60 & 48 & 1.35 & 10 & 2.77 & 0.96 & 0.347 & 0.35 \\
18 & 2.20 & 7 & 1.04 & 1.70 & 1.637 & 1.64 & 49 & 1.30 & 10 & 3.06 & 1.68 & 0.550 & 0.55 \\
19 & 2.18 & 7 & 1.08 & 2.00 & 1.852 & 1.85 & 50 & 1.29 & 10 & 3.15 & 2.15 & 0.685 & 0.69 \\
20 & 2.17 & 7 & 1.08 & 2.49 & 2.304 & 2.30 & 51 & 1.29 & 10 & 3.15 & 1.45 & 0.460 & 0.46 \\
21 & 2.15 & 7 & 1.12 & 1.70 & 1.514 & 1.51 & 52 & 1.20 & 10 & 3.89 & 1.74 & 0.447 & 0.45 \\
22 & 2.14 & 7 & 1.14 & 1.97 & 1.731 & 1.73 & 53 & 1.16 & 11 & 3.43 & 1.57 & 0.459 & 0.46 \\
23 & 2.08 & 7 & 1.23 & 1.19 & 0.965 & 0.96 & 54 & 1.15 & 11 & 3.54 & 1.63 & 0.460 & 0.46 \\
24 & 2.03 & 8 & 1.09 & 2.19 & 2.005 & 2.00 & 55 & 1.13 & 11 & 3.73 & 1.65 & 0.443 & 0.44 \\
25 & 1.98 & 8 & 1.18 & 1.61 & 1.362 & 1.36 & 56 & 1.12 & 11 & 3.78 & 1.79 & 0.474 & 0.47 \\
26 & 1.89 & 8 & 1.36 & 2.00 & 1.466 & 1.47 & 57 & 1.11 & 11 & 3.85 & 2.19 & 0.568 & 0.57 \\
27 & 1.85 & 8 & 1.46 & 2.01 & 1.378 & 1.38 & 58 & 1.07 & 11 & 4.29 & 1.65 & 0.385 & 0.39 \\
28 & 1.81 & 8 & 1.56 & 1.49 & 0.954 & 0.95 & 59 & 1.06 & 11 & 4.43 & 1.50 & 0.339 & 0.34 \\
29 & 1.72 & 8 & 1.80 & 1.88 & 1.044 & 1.04 & 60 & 1.06 & 11 & 4.49 & 2.35 & 0.523 & 0.52 \\
30 & 1.71 & 8 & 1.84 & 1.47 & 0.801 & 0.80 & 61 & 1.03 & 11 & 4.82 & 2.16 & 0.449 & 0.45 \\
31 & 1.71 & 8 & 1.85 & 1.99 & 1.077 & 1.08 & 62 & 1.01 & 12 & 4.37 & 1.32 & 0.303 & 0.30 \\
\bottomrule
\end{tabular}

\end{center}

\clearpage
\paragraph{Development along the central airways}
The lumen is narrowest at the inlet: the inlet cap has an area of 59.7\,mm$^2$ and the first section normal to the
axis, 76\,mm above the carina, 74\,mm$^2$. Over the next 45\,mm the lumen widens to 145\,mm$^2$,
the mean axial velocity falls from 1.36 to 0.69\,m\,s$^{-1}$, and the area-averaged pressure rises from 3.43 to
3.96\,Pa at S1 (Fig.~\ref{fig:cfd-profiles}a--c). The jet does not spread to fill the widened lumen: its peak axial
velocity falls only from 2.95 to 2.11\,m\,s$^{-1}$ over the trachea, and it runs along one side of the lumen
(Fig.~\ref{fig:cfd-profiles}e) while reverse axial flow occupies up to 31\,\% of the tracheal cross-section beside it.
The $-x$ main bronchus divides about 20\,mm from the carina, so its profile is shown only up to that point. In both
main bronchi the secondary flow generated at the carina decays with distance: in the $+x$ bronchus, the ratio of mean
transverse to mean axial velocity falls from 0.38 at S3 to 0.12 at 40\,mm (Fig.~\ref{fig:cfd-profiles}d).

\begin{center}
    \includegraphics[width=0.88\textwidth]{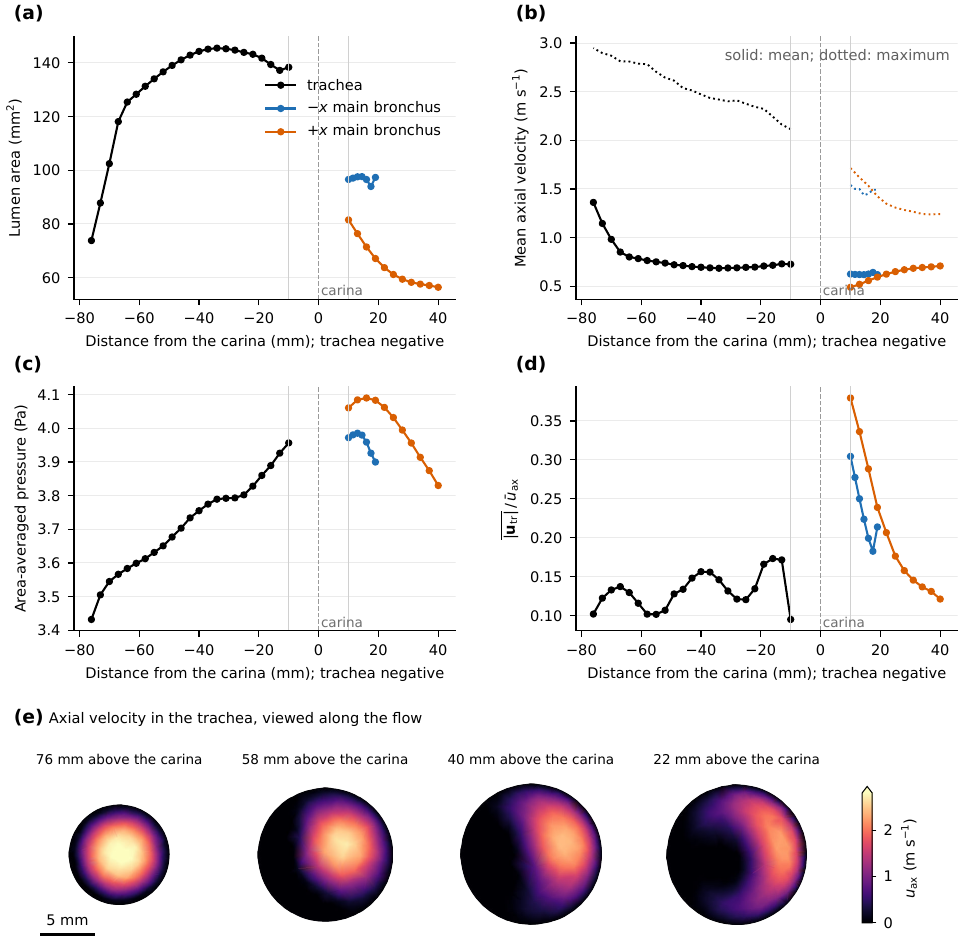}
    \captionof{figure}{Flow development along the trachea and main bronchi, on sections normal to the local airway
    axis, every 3\,mm along the trachea and the $+x$ main bronchus and every 1.5\,mm along the $-x$ main bronchus up to
    its first division. Distances are measured from the carina along each axis, negative in the trachea; thin vertical
    lines mark S1--S3. Quantities are defined in Eq.~\eqref{eq:s-profiles}.
    (a)~Lumen area. (b)~Mean (solid) and maximum (dotted) axial velocity. (c)~Area-averaged static pressure.
    (d)~Ratio of the mean transverse to the mean axial velocity, a measure of secondary flow.
    (e)~Axial velocity at four tracheal sections, viewed along the flow, on a common scale.}
    \label{fig:cfd-profiles}
\end{center}

\clearpage
\paragraph{Outlets}
Outlet flow share scales with outlet diameter as $D_k^{2.24}$ ($R^2=0.94$; Fig.~\ref{fig:cfd-outlets}b), and the
ten largest outlets together carry 45\,\% of the inflow. Of the pressure drop from the inlet to the distal end of each
subtree, between 26 and 71\,\% (median 50\,\%) arises within the reconstructed airways, the rest within the modelled
subtree (Fig.~\ref{fig:cfd-outlets}c). The complete per-outlet results are listed in Table~\ref{tab:cfd-outlets}.

\begin{center}
    \includegraphics[width=0.86\textwidth]{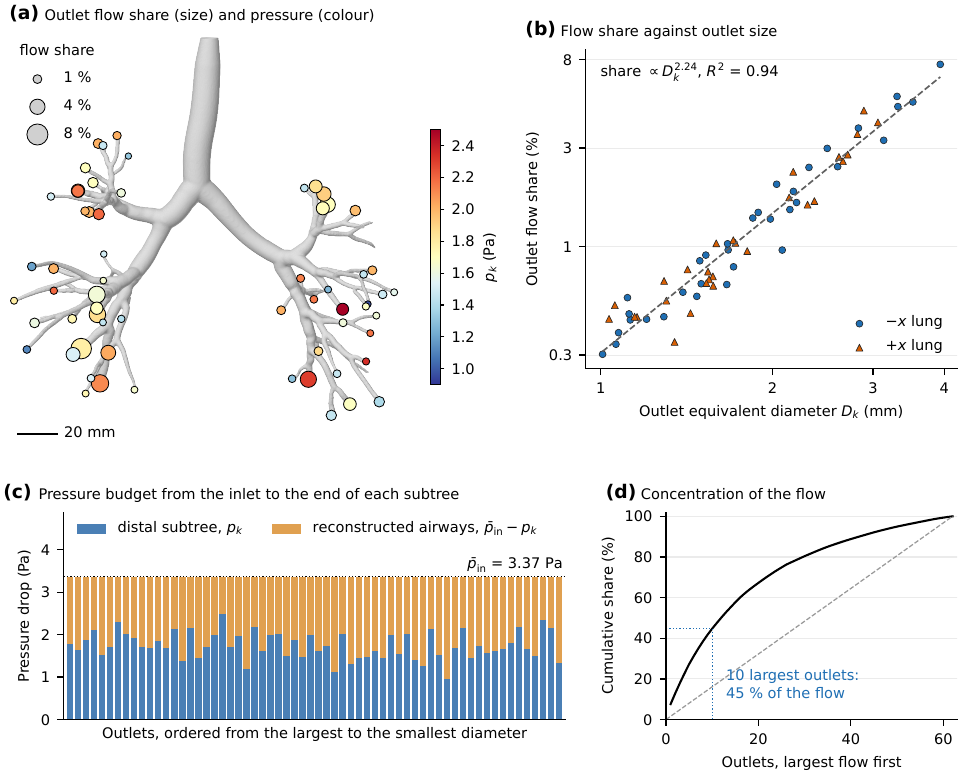}
    \captionof{figure}{Distribution of flow and pressure over the 62 outlets.
    (a)~Outlet flow share (marker area) and outlet pressure $p_k$ (colour).
    (b)~Flow share against outlet equivalent diameter, with a power-law fit.
    (c)~Pressure budget for each outlet: the drop from the mean inlet pressure to the outlet, within the
    reconstructed airways, stacked on the drop across the modelled distal subtree, $p_k=R_kQ_k$ (Eq.~\eqref{eq:s-budget}).
    (d)~Cumulative flow share, outlets ordered by decreasing flow; the dashed line is an equal split.}
    \label{fig:cfd-outlets}
\end{center}

\vspace{4mm}
\section*{Supplementary Note S1: Development observations}
\phantomsection\label{sec:supp-note}

\paragraph{Defects found during validation}
All defects found and corrected while developing the validation lay in surface clean-up (port opening,
degenerate-triangle removal, non-manifold repair and component selection) rather than in graph extraction, the
shared-port construction or the implicit field. (i)~Port opening originally assigned each triangle to its
Euclidean-nearest port, irrespective of branch; a large-radius port near an unrelated branch could sever the neck
joining a sub-tree to the trunk, after which a largest-component filter discarded the sub-tree (one case lost a
16-outlet lobe). Restricting removal to the port's own branch and its immediate parent resolved this, and in four
further cases blocked many triangles without changing the outcome. (ii)~A component filter implemented as a
re-triangulation round trip perturbed the mesh even when it removed nothing, seeding fragmentation in three cases;
direct cell deletion removed the problem. (iii)~Port-to-loop assignment without a rejection rule redistributed
neighbouring matches when one loop was missing, reporting errors above 100\,mm on ports whose geometry was sound; the
tolerance-based rejection of Section~\ref{sec:acceptance} prevents this. Each defect was invisible in the final output
and was detected only by inspecting intermediate stages.

\paragraph{Length-to-diameter conventions}
Measured as arc length over the full graph edge, the pooled length-to-diameter ratio was 8.99. Using the chord between
successive branch points reduced it to 7.63. Terminal pruning had also merged branches: removing one child of a
bifurcation leaves a pass-through node, which was collapsed so that two anatomical branches formed one edge. Such edges
were 19.7\,\% of all branches and had a median ratio of 15.3 against 5.4 for unmerged branches. Keeping former
bifurcations as segment boundaries, together with chord lengths, gives the values of
Tables~\ref{tab:morphometry} and~\ref{tab:ld-generation}; because a merged edge's parent had been its grandparent, it
also raised the child-to-parent diameter ratio.

\end{document}